\documentclass[prb,twocolumn,showpacs,amsmath,amssymb,superscriptaddress,floatfix]{revtex4}

\usepackage[english]{babel}
\usepackage{amsfonts}
\usepackage{graphicx}
\usepackage{times}

\usepackage{color}
\definecolor{nred} {RGB}{224,0,0}
\definecolor{nblue} {RGB}{28,130,185}
\definecolor{dgreen} {RGB}{78,138,21}

\begin{document}

\title{Spectral Function  of an  Electron Coupled to  Hard Core Bosons}

\author{J. \surname{Bon\v ca}}
\affiliation{Faculty of Mathematics and Physics, University of Ljubljana, 1000
Ljubljana, Slovenia}
\affiliation{J. Stefan Institute, 1000 Ljubljana, Slovenia}



\date{\today}
\begin{abstract}
We have computed  static and dynamic properties of an electron coupled to hard--core--boson (HCB)  degrees of freedom in one spacial dimension. 
The polaron, an electron dressed with HCB excitations,  remains light even in the strong coupling limit as  its effective mass remains  of the order of the free electron mass. This result  is in a sharp contrast to the Holstein model where the electron effective mass increases exponentially with the electron--phonon coupling. HCB degrees of freedom mediate the  attractive potential between two electrons that form a bound singlet bipolaron state  at any non--zero coupling strength.  In the low--frequency regime of the electron spectral function we observe a quasi--particle (QP) band that is  separated from the continuum of states only in the central part of the Brillouin zone. 
The quasiparticle weight  approaches zero as the QP band  enters  the continuum  where it obtains a finite lifetime.  At finite temperature an electron can annihilate thermally excited HCB's. Such thermally activated processes   lead to a buildup of the spectral weight below the QP band. While the investigated model bears a resemblance with the Holstein model, we  point out many important differences that originate from the binary  HCB excitation  spectrum, which in turn mimics  spin-$1\over 2$ degrees of freedom. 
\end{abstract}  

\maketitle

\setcounter{figure}{0}

\section{Introduction}

We investigate static, dynamic and thermodynamic properties of an electron coupled to hard--core--boson (HCB) degrees of freedom. The model under  the investigation resembles a well  known Holstein model (HM),\cite{holstein59} which has been a subject of extensive research.\cite{engelsberg63,bonca1,bonca4,Ranninger1992,marsiglio93,alexandrov94,fehske1997,fehske2000,osor2002,osor2004,osor2006,filippis2005,prokofev98,mishchenkoSELF,berciu07a,hohen2003,berciu07b,goodvin,Bonca_2019} Starting from the HM,  the HCB model (HCBM) is obtained by replacing each oscillator with its infinite degrees of freedom by a HCB representing a state  that can be either occupied or un--occupied.  Despite a significant reduction of inelastic degrees of freedom  in comparison to the HM, the statistics of neighboring energy level
spacings in HCBM remains  well described by the Wigner-Dyson distribution,\cite{Vidmar_2019} 
characteristic of non--integrable  ergodic quantum models. The reduction of HM to HCBM has further important consequences: HCBM possess a limited  energy spectrum which further facilitates finite--temperature studies, reduction of inelastic degrees of freedom allows studies of larger--size systems approaching closer the thermodynamic limit. 

An important motivation  for the research presented in this manuscript originates from the research of high--temperature superconductivity   where the quest  for the origin of the attractive interaction between charge carriers is still active. Among the fundamental open questions is whether the attractive interaction is based on lattice degrees of freedom or is it due to the strong Coulomb interaction that generates  the spin exchange coupling.  Since the  binary spectrum of a HCBs closely resembles a spin $1/2$ degree of freedom, HCBM can be used to simulate properties of a spin-polaron\cite{Brinkman_1970,Trugman_1988,Shraiman_1988}, {\it i.e.} electron, interacting with   spin degrees of freedom thus bridging the gap  between research of  lattice and spin models.

So far physical realization of systems with HCB's is limited to ultracold atoms in optical lattices.\cite{Vidmar_2015,Bloch_2013} Nevertheless, models where electrons are coupled to HCB's may carry potential relevance to microscopic mechanisms of superconductivity. As we will show in this paper, the effective polaron mass remains small even in the limit of very strong electron-HCB coupling strength.  We also demonstrate  that HCB's can mediate   the necessary attractive potential for the formation of light bipolarons.

\section{Model and method}

We analyze a model with a single electron in a one--dimensional chain of size $L$ with periodic  boundary conditions coupled to HCB  degrees of freedom
\begin{eqnarray} \label{hol}
\vspace*{-0.0cm}
H &=& -t_0\sum_{j}(c^\dagger_{j} c_{j+1} +\mathrm{H.c.})-{g} \sum_{j} \hat n_{j} (b_{j}^\dagger + b_{j}) \nonumber \\
 &+&  \omega_0\sum_{j} b_{j}^\dagger  b_{j} 
 , \label{ham}
\end{eqnarray}
where $c^\dagger_{j}$ and $b^\dagger_{j}$ are electron and HCB creation operators at site $j$, respectively, and $\hat n _{j} = c^\dagger_{j}c_{j}$ represents the  electron density operator. HCB's are defined via the following commutation relation $[b_i,b_j^\dagger]=\delta_{i,j}(1-2b_i^\dagger b_i)$.
Parameter $g$ measures the strength of coupling between the electron and HCB, $\omega_0$ denotes a dispersionless  optical HCB  frequency and $t_0$ nearest--neighbor hopping amplitude. From here on we set $t_0=1$. 

In most calculations we have used  full translationally invariant Hilbert space spanning $N_\mathrm{st}=2^L$ basis states. In case of zero-$T$ calculations we have implemented standard Lanczos\cite{lanczos} technique.  To determine static and dynamic properties of the model at finite-T we have implemented the Finite Temperature Lanczos Method (FTLM) as described in Refs.~\cite{jaklic2000,prelovsek_book} where it has been  shown that  static thermodynamic properties of an operator $A$ can be evaluated via sampling over random states $\vert r\rangle$ defined in a subspace with one electron and multiple HCB degrees of freedom,
\begin{eqnarray}
\langle A \rangle_T &=&  {\cal Z}^{-1} \sum_{r=1}^R  \sum_{j=1}^M e^{-\beta \epsilon_j} \langle  r \vert \psi_j \rangle \langle \psi_j \vert  A \vert  r\rangle, \label{ave}\\ 
{\cal Z} &=& \sum_{r=1}^R  \sum_{j=1}^M e^{-\beta \epsilon_j} \vert \langle  r \vert \psi_j \rangle \vert^2, \nonumber
\end{eqnarray}
where $\vert \psi_j\rangle$ and $\epsilon_j$ are Lanczos wave-functions and corresponding energies, respectively, in the sub-space with one electron, $\beta=1/T$,  and $\cal Z$ is the partition function. Furthermore, $R$ represents the number of different random states,  and $M$ is the number of Lanczos iterations.

The main goal of this work is to analyze the   single-electron spectral function, corresponding to electron addition,  as obtained  via the corresponding retarded Green's function 
\begin{equation}
A(\omega,k)=-\pi^{-1}\lim_{\eta\to 0^+} {\cal G}^R( \omega + i \eta,k),\label{akom}
\end{equation}
where at finite-$T$ ${\cal G}^R( \omega + i \eta,k)$ is obtained via the Gibbs ensemble 
\begin{equation}
{\cal G}^R(\omega,k) = {\cal Z}^{-1} \int_0^\infty dt~e^{i\omega t}\sum_n e^{-\beta \epsilon_n^0} \langle \phi_n^0\vert c_k(t)c_k^\dagger (0)\vert\phi_n^0\rangle,\label{green}
\end{equation}
where $\phi_n^0$ are multi HCB eigenstates of  the Hamiltonian in Eq.~\ref{ham} with no electron present in the system while $\epsilon_n^0$ are  corresponding energies and $c_k = 1/\sqrt{L} \sum_{j=1}^L \exp(ikj) c_j$. We finally  take advantage of the FTLM method,\cite{jaklic2000}  and replace the  trace over a complete set of states  states $\vert \phi_n^0\rangle$  by the summation over random states $\vert r^0\rangle=\sum_{n=1}^{N_0} \alpha_n \vert\phi_n^0\rangle$, where $\alpha_n$ are distributed randomly, and
\begin{eqnarray}
A(\omega,k) = {\cal Z}^{-1}&&  \sum_{r=1}^R\sum_{n=1}^{N_{0}}\sum_{j=1}^M e^{-\beta \epsilon_n^0} \langle r^0\vert \phi_n^0\rangle 
\langle \phi_n^0\vert c_k \vert \psi_j \rangle \nonumber \\ 
&\langle&\psi_j \vert c_k^\dagger \vert r^0\rangle\delta(\omega-\epsilon_j+\epsilon_n^0),\label{akom}
\end{eqnarray}
where $\vert \psi_j\rangle$ and $\epsilon_j$ are Lanczos wave-functions and corresponding energies, respectively, in the sub-space with one electron. Lanczos states are generated starting from  states $ c_k^\dagger \vert r^0\rangle$. Furthermore, $R\sim 100$ represents the number of different random states, $N_0=2^L$ the size of the Hilbert space in the zero-electron subspace and $M$ is the number of Lanczos iterations. We have used typically $M=50$ Lanczos iterations to obtain static properties at finite-$T$ and $M=500$ for dynamic quantities combined with the Gram-Schmidt orthogonalization procedure to avoid spurious non-orthogonal states that appear that appear at  large values of $M$. The Lorentzian form of the delta functions with the half width at half maximum (HWHM) $\eta $ was used  for graphic  representations of $A(k,\omega)$. 

While our calculations were limited to rather small system sizes $L=16$, we have expanded our calculations from using only periodic boundary conditions towards the  twisted boundary conditions,\cite{shastry,Poilblanc_1991,Bonca_2003,Bonca_2019} that are equivalent to the introduction of the  magnetic flux penetrating the ring.  In this approach the kinetic energy term in Eq.~(1) is transformed to
$
H_\mathrm{kin} = -t_0\sum_{j}(c^\dagger_{j} c_{j+1} e^{i\theta}+\mathrm{H.c.}), 
$
where  $\theta$  represents a magnetic flux that penetrates the ring  $\phi_m=\theta L/2\pi$ in units of $h/e_0$.  Discrete $k-$ points $k_n=2\pi n/L$ can thus  be  connected  by choosing $\theta \in [0,2\pi/L]$. 

Further details concerning the construction of the translationally invariant basis states and comparison of the selected ground state properties between a small basis set and a full basis  can be obtained in the Appendix A. 
We show in the  Appendix B that  finite-size effects seem to be well under control in all temperature  regimes by comparing $A(k,\omega)$ computed on two different system sizes and different $T$.  In the Subsection  E we also compute the lowest frequency moments of $A(k,\omega)$ and compare them with analytical results that are free of finite--size effects, which in turn  renders further insight into the applicability of FTLM to the model under the investigation. 

\section{Results}

\subsection{Zero-$T$ properties}

\begin{figure}[!tbh]
\includegraphics[width=0.8\columnwidth]{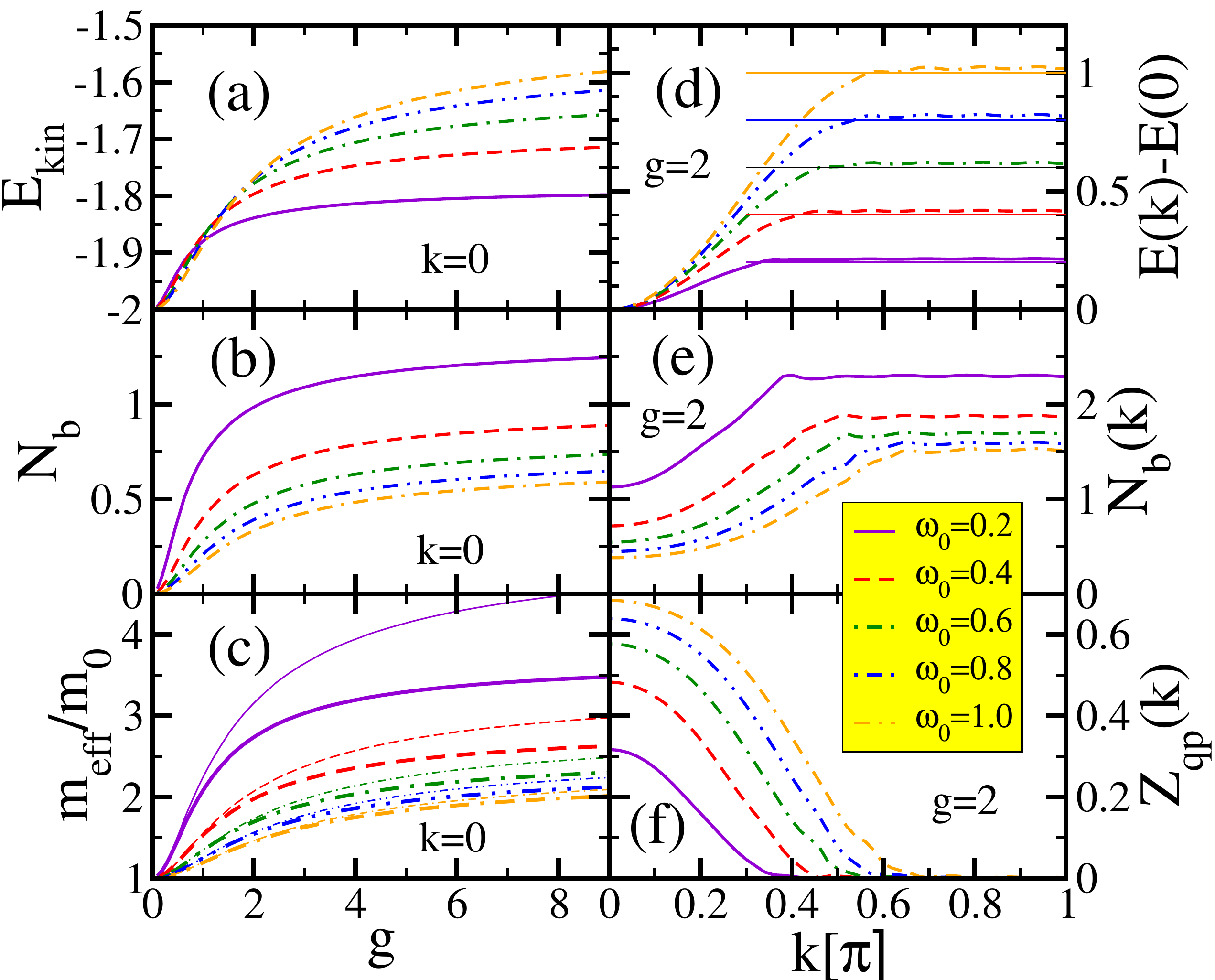}
\caption{ Ground-state properties computed using full basis on a ring with $L=16$ taking into account the  full translational symmetry: a) the kinetic energy $E_\mathrm{kin} = \langle \psi_k\vert -t_0\sum_j c_j^\dagger c_{j+1} + \mathrm{H.c.}\vert \psi_k \rangle$ vs. $g$ at different $\omega_0$,  b) the total number of HCBs $N_b =\langle \psi_k\vert \sum_j b_j^\dagger b_j\vert \psi_k\rangle $ and c) the effective mass $m_\mathrm{eff}=(\partial ^2 E(k)/\partial k^2\vert_{k=0}  )^{-1}$ compared to the free electron mass $m^0=1/2t_0$ where $E(k)=\langle \psi_k \vert H \vert \psi_k \rangle$, where $\vert \psi_k\rangle$ is the ground state polaron wave function with momentum $k$.  Thin lines represent the inverse of the quasiparticle weight  defined  as $Z_\mathrm{qp}(k)=\vert \langle \psi_k\vert c_k^\dagger \vert \emptyset\rangle\vert^2$ at $k=0$ where $\vert \emptyset \rangle$  is the state with no electron and no HCB excitations. In d) through f) we show $k-$ dependent properties at fixed coupling $g$ of the total energy $E(k)$ in d), $N_b(k)$ in e) and in f) the quasiparticle weight $Z_\mathrm{qp}(k)$. Horizontal lines in d) indicated values of $\omega_0$. }
\label{fig1}
\end{figure}

In Figs.~\ref{fig1}(a) through (c)  we show some characteristic ground state properties of the model at $k=0$. With increasing the electron-boson coupling $g$  the kinetic energy $E_\mathrm{kin}$ increases but shows a tendency towards saturation  at larger $g$, as seen in Fig.~\ref{fig1}(a). $E_\mathrm{kin}$ also increases with increasing $\omega_0$ at fixed $g$ which stands    in contrast to a decrease   of the total number of HCB's $N_b$ with increasing $\omega_0$, seen in Fig.~\ref{fig1}(b).  Further elaborating  on   results at fixed $g$: at smaller $\omega_0$ the electron is dressed up with a larger number of HCB excitations than at larger values of $\omega_0$, nevertheless, its kinetic energy is less affected by the presence by the HCB cloud at smaller $\omega_0$. It is furthermore instructive to compare the  slow increase of $N_b$ with increasing $g$  to the   increase of the number of phonons  in the strong coupling limit in the HM where $N_\mathrm{ph}= (g/\omega_0)^2$. This profound  difference between the two models is trivially explained since the maximal on--site occupancy of HCB's is limited to 1. In Fig.~\ref{fig1}(c) we display   the  effective polaron mass $m_\mathrm{eff}$.  In the small--$g$ limit the perturbation theory  can be applied   and results in the following polaron dispersion relation:
\begin{equation}
E(k)=\epsilon(k) +{1\over L}\sum_q {g^2\over \epsilon(k)-\epsilon(k+q)-\omega_0},
\end{equation}
where $\epsilon(k)=-t_0\cos(k)$ is the free electron dispersion relation. In this limit  $m_\mathrm{eff}/m_0-1 \propto g^2$ just as in the HM, while at larger $g$  it displays  a slow sub--logarithmic increase with $g$. 
Here, the difference with the HM is even more pronounced since it is well known that in  the strong coupling limit of the HM $m_\mathrm{eff}/m_0=\exp[(g/\omega_0)^2]$. In the case of the HM $m_\mathrm{eff}$ can be  determined from  $Z_\mathrm{qp}(k=0)^{-1} = m_\mathrm{eff}/m_0$, as shown in Ref.~\cite{fehske2000} In Fig.~\ref{fig1}(c) we show along    $m_\mathrm{eff}/m_0$ also $Z_\mathrm{qp}^{-1}$. The agreement between both quantities is  good for large $\omega_0$ at any $g$, but starts to deviate strongly at small $\omega_0$.

\begin{figure}[!tbh]
\includegraphics[width=0.8\columnwidth]{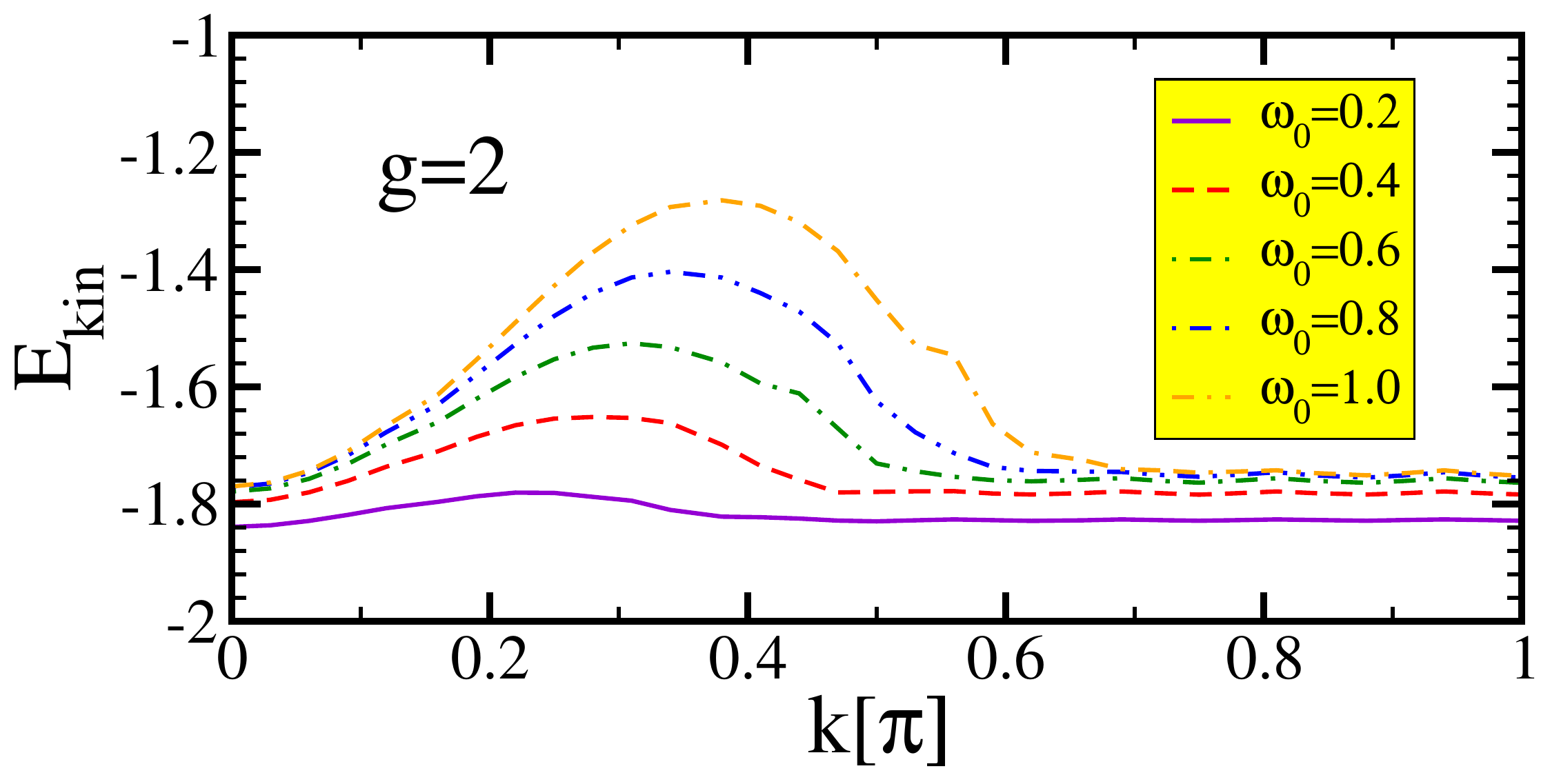}
\caption{ Electron kinetic energy $E_\mathrm{kin}$  vs. $k$ at different $\omega_0$.   }
\label{Fig2}
\end{figure}

In Figs.~\ref{fig1}(d) through (f)  we show $k$--dependent properties of the electron coupled to HCB's at fixed $g=2$ and various $\omega_0$. We start with the discussion of the dispersion relation representing the polaron energy  $E(k)=\langle \psi_k \vert H \vert \psi_k \rangle$, where $\vert \psi_k\rangle$ is the ground state polaron wave function with momentum $k$,  shifted by $E(k=0)$ which facilitates a quantitative  comparison between results corresponding to different values of $\omega_0$, as shown in Fig.~\ref{fig1}(d). Common to all cases is a quadratic increase at small $k$ followed by an abrupt flattening of the band at a value of $k_0$ corresponding to  $E(k_0)-E(0)=\omega_0$. It seems as if the polaron band, also known as the quasi--particle (QP) band  intercepts  the  continuum of states that starts at the energy $\omega_0$ above $E(0)$.  This is in a sharp contrast to the Holstein polaron case where the  QP band remains separated from the continuum in the whole Brillouin zone. A similar $k$--dependence is observed in the mean HCB number $N_b(k)$, Fig.~\ref{fig1}(e).  The   QP weight $Z_\mathrm{qp}(k)$ provides information about the electronic character of the polaronic state.  In Figs.~\ref{fig1}(f)  we observe  a disappearance of $Z_\mathrm{qp}$ at  $k_0$ where the QP band enters the continuum.  This result suggests  that the ground state $\vert\psi_k\rangle$ for $k>k_0$  contains nearly zero free electron contribution of a state $c_k^\dagger\vert\emptyset\rangle$. Instead, $\vert\psi_k\rangle$ is composed of a polaron in the state $k=0$ that contains a significant amount of the free--electron wavefunction  $c_{k=0}^\dagger\vert \emptyset\rangle$,  and  an extra HCB excitation in the state $k$. This is  further  consistent with results presented in  Fig.\ref{Fig2} where  we show  the electron kinetic energy $E_\mathrm{kin}(k)=\langle \psi_k \vert H_\mathrm{kin} \vert \psi_k \rangle$. Naively, one would expect  $E_\mathrm{kin}(k)$ to resemble a renormalized free--electron dispersion relation.  In contrast, $E_\mathrm{kin}(k)$ shows a non--monotonic behavior whereby at larger $k>k_0$ it again reaches its value at zero momentum, {\it i.e.} $E_\mathrm{kin}(k=0)\sim E_\mathrm{kin}(k>k_0)$, consistent with the above hypothesis.  


\begin{figure}[!tbh]
\includegraphics[width=1.0\columnwidth]{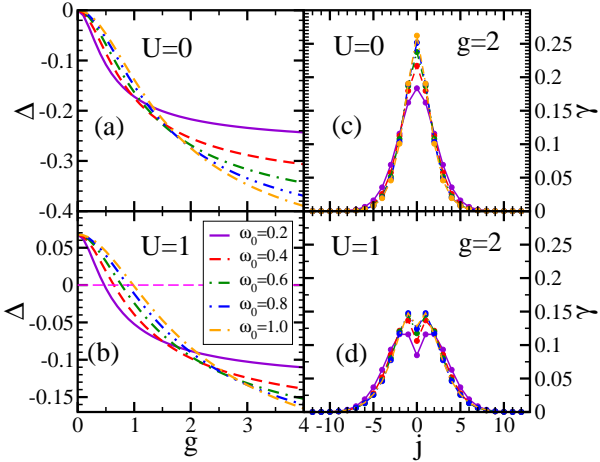}
\caption{ $\Delta$ vs. $g$ for different values of $\omega_0$ as shown in the legends for $U=0$ and 1 in (a) and (b), respectively. In (c) and (d) we  present $\gamma(j)$ at fixed $g=2$ while the other parameters of the model are the same as in (a) and (b).   All results were computed on a system with 13 sites using periodic boundary conditions. 
}
\label{Fig3}
\end{figure}

Before switching to the description of finite--$T$ properties of the HCBM we show that  the electron--HCB interaction leads to a formation of bound pairs of polarons or bipolarons.  To test this assumption  we investigate a system with  two electrons coupled with HCB's in the presence of a Coulomb on-site interaction $U$
\begin{eqnarray} \label{hol}
\vspace*{-0.0cm}
H &=& -t_0\sum_{j,\sigma\in [\uparrow,\downarrow]}(c^\dagger_{j,\sigma} c_{j+1,\sigma} +\mathrm{H.c.})-{g} \sum_{j} \hat n_{j} (b_{j}^\dagger + b_{j}) \nonumber \\
 &+& U\sum_{j}n_{j\uparrow}n_{j\downarrow}+ \omega_0\sum_{j} b_{j}^\dagger  b_{j}, 
 ,\label{hambi}
\end{eqnarray}
where $n_j = n_{j\uparrow}+n_{j\downarrow}$ and $n_{j,\sigma}=c_{j,\sigma}^\dagger c_{j,\sigma}$.
We have computed the binding energy of a bipolaron defined as $\Delta = E_2 - 2E_1$ where $E_2$ is the ground state energy of two electrons with opposite spin orientation  and $E_1$ is the ground state energy of a system with one electron.  In  Fig.~\ref{Fig3} we present $\Delta$ vs. $g$ for different values of the HCB frequency $\omega_0$. Results show that at $U=0$ the bipolaron remains bound irrespective of the coupling strength as well as $\omega_0$ while at finite $U=1$ there exists a critical value of  $g_c$. This result is consistent with $\Delta$ computed in the  atomic limit, {\it i.e.} at $t_0=0$ where we obtain the following  expression for the binding energy:
\begin{equation}
\Delta = \sqrt{4g^2+\omega_0^2}-{1\over 2}\sqrt{16g^2+\omega_0^2}-\omega_0/2 + U,
\label{bhcb}
 \end{equation}
which leads to a $\Delta<0$ for $U=0$ at arbitrary $|g|>0$.

To get further insight into the shape of the bipolaron we also present in Fig.~\ref{Fig3}  the density--density correlation function 
\begin{equation}
\gamma(j)=\frac{\sum_{i} \langle  \psi_0 | n_i n_{i+j}| \psi_0  \rangle}{\sum_{i,l} \langle  \psi_0 | n_i n_{i+l}| \psi_0  \rangle },
\label{gamma}
\end{equation} 
defined so as $\sum_j\gamma(j)=1$. The $\Delta<0$ regime is reflected in  an exponentially decaying form of $\gamma(j)$. At finite--$U$ the bound state is formed from two electrons positioned predominantly on neighboring sites. This results bears a similarity to a $S1$ bipolaron in the Holstein--Hubbard model\cite{fehske1995,magna1997,bonca2000,alexandrov07}. 

\begin{figure}[!tbh]
\includegraphics[width=0.8\columnwidth]{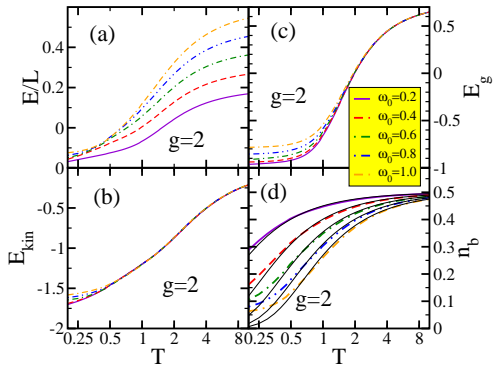}
\caption{ Finite-$T$ properties  of selected expectation values computed on a system with $L=16$ presented on a semi--log plot.  In a) we present the energy of the system $E(T)=\langle H\rangle_T$ per site $L$ where the expectation value is taken using FTLM method, as described in Eq.~\ref{ave}; similarly,  in b) we present $E_\mathrm{kin}(T)$, in c) we display the electron--HCB coupling part of the Hamiltonian in Eq.~\ref{ham}:  $E_g=-\langle {g} \sum_{j} \hat n_{j} (b_{j}^\dagger + b_{j})\rangle_T$ and in d)  the number of HCB per site $n_b=N_b/L$ along with the HCB thermal distribution functions  $n_b^0={1\over \exp[\beta\omega_0]+1}$, shown using thin black lines. }
\label{Fig4}
\end{figure}

\subsection{Finite-$T$ properties}
In Figs.~\ref{Fig4} we present selected thermodynamic properties of the model. The total energy of the system increases with $T$     as well as with $\omega_0$, see Fig.~\ref{Fig4}(a).  This is in part due to  the thermal  increase of the average HCB site occupation number $n_b$. The other contribution comes from the increase of the kinetic energy $E_\mathrm{kin}$, shown in Fig.~\ref{Fig4}(b),  that with increasing $T$ approaches its high-$T$ limit $E_\mathrm{kin}=0$. Unexpectedly, $E_\mathrm{kin}$ shows very weak  dependence on the HCB frequency $\omega_0$ except at very small $T$.  Likewise, the electron-HCB coupling part of the Hamiltonian, $E_g$ as shown in Fig.~\ref{Fig4}(c)   increases with $T$ and even changes sign, also displays  very little  dependence on $\omega_0$ above $T\gtrsim 1$. The  HCB occupancy per site  $n_b$  closely follows the Fermi--Dirac--like distribution, characteristic for HCBs $n_b^0\sim 1/[\exp(\beta\omega_0)+1]$, as shown  in Fig.~\ref{Fig4}(d) by thin black lines.  In the high-$T$ limit, {\it i.e.} for for $T>>\omega_0$,  numerical results clearly approach $n_b\to 1/2$. 

\begin{figure}[!tbh]
\includegraphics[width=0.8\columnwidth]{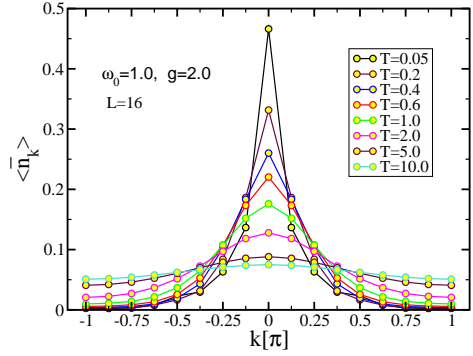}
\caption{The single particle density matrix $\bar n_k$ vs. $T$ computed using $\omega_0=1$,  $g=2$ and $L=16$. 
 }
\label{Fig5}
\end{figure}

In Fig.~\ref{Fig5} we present the single  particle density matrix $\bar n_k$ defined as $\langle c_k^\dagger c_k\rangle_T$ where the thermal average is taken in the sub-space with one electron. In addition, $\bar n_k$ represents the sum--rule of the electron--removal spectral function. In the case of the free electron, {\it i.e.} at $g=0$,  $\bar n_k(T=0)=\delta_{k,0}$, while at finite $g>0$ $\bar n_k(T=0)$ remains centered around $k=0$ but obtains a finite width in the momentum space. With increasing $T$ its width  increases and in the $T\to\infty$ limit approaches a constant $\bar n_k = 1/L$. We should also stress that besides the  sum--rule $\sum_k \bar n_k=1$ there exist another slightly less obvious relation $-2t_0\sum_k \cos(k) \bar n_k=E_\mathrm{kin}(T)$.
\begin{figure}[!tbh]
\includegraphics[width=1.0\columnwidth]{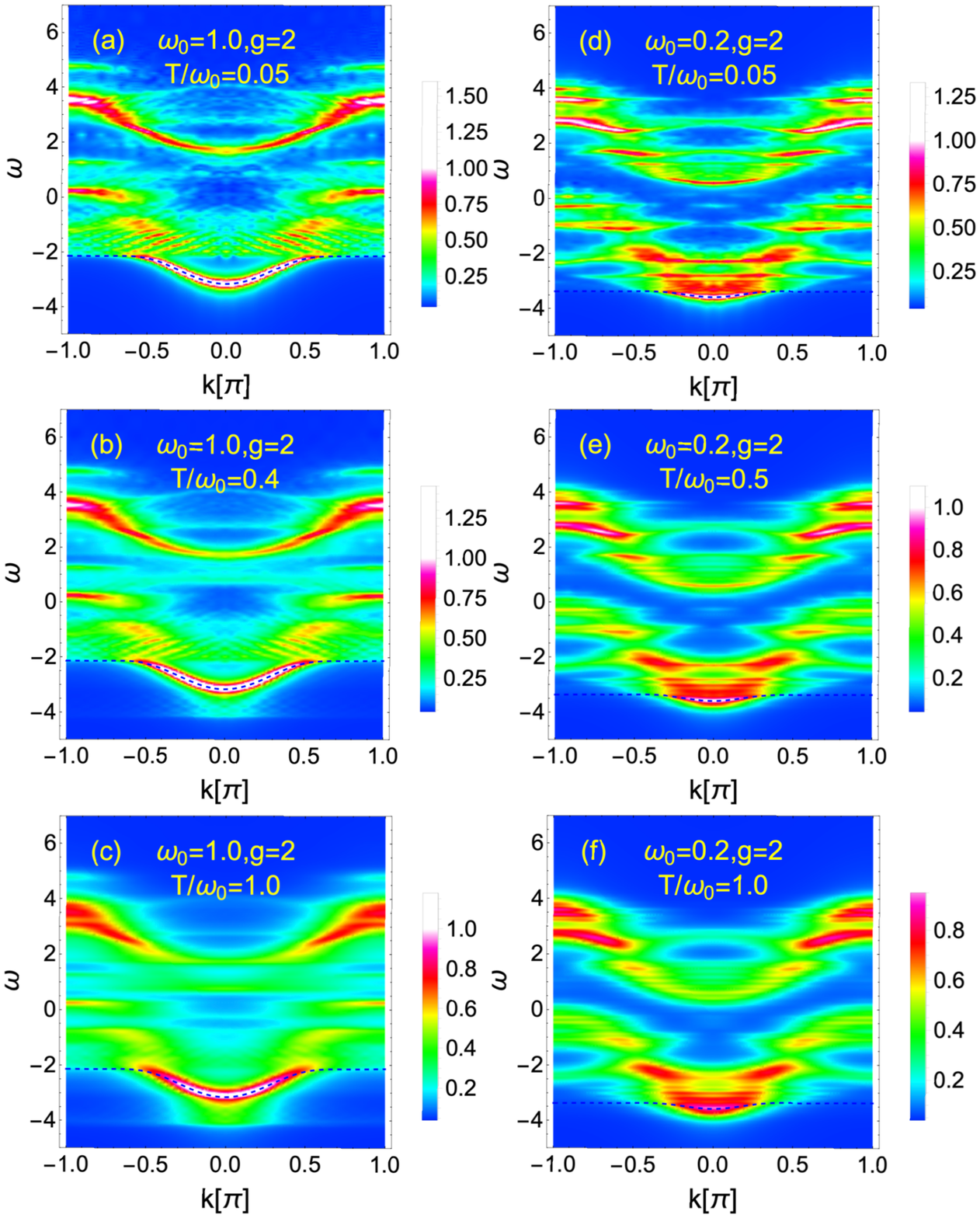}
\caption{$A(\omega,k)$ for $g=2$ vs. different values of $T/\omega_0$ presenting two sets of results in parallel for  $\omega_0=1$ in (a) through (c) and 
$\omega_0=0.2$ in (d) through (f). The same size of the system $L=16$ was used as in Fig.~\ref{fig1}. In addition we have used twisted boundary conditions to compute  $A(\omega,k)$ at 25 equally spaced  $k-$ points in the interval $k\in [0,\pi]$ with increments of $\Delta k = \pi/24$. Moreover, in all figures from (a) to (f) the same color coding was used to enable direct comparison between different cases. In all figures we also display the  polaron dispersion relation at zero-$T$  $E(k)$   using a dashed line as a guide to the eye. Lorentzian broadening $\eta=0.05$ was used in all cases.}
\label{Fig6}
\end{figure}

\subsection{Spectral functions $A(k,\omega)$}

In Figs.~\ref{Fig6} we present density plots representing $A(\omega,k)$ in the entire Brillouin zone (BZ) at fixed coupling $g=2$ and two different sets of $\omega_0=1$ and 0.2.  For a more quantitative analysis   we also display with dashed lines  the corresponding ground--state polaron energy $E(k)=\langle \psi_k \vert H \vert \psi_k \rangle$, where $\vert \psi_k\rangle$ is the ground state polaron wave function with momentum $k$. We focus  first on $\omega_0=1$  and  small $T/\omega_0=0.05$  where we clearly observe in the low--$\omega$ regime the QP  band only  in a limited part of the Brillouin zone,  $k\in[-0.6\pi,0.6\pi]$. The spectral weight of the QP band is given by $Z_\mathrm{qp}(k)$. Outside this interval the QP band enters the continuum of states and   $Z_\mathrm{qp}$ approaches  zero, also consistent with    Figs.~\ref{fig1}(d) and (f).  

 In the  Appendix \ref{B} we show that only a partial separation of the QP band from the continuum is a consequence of the HCB commutation relation.  In a model where the Hilbert space of HCBM is expanded  to contain up to two phonon degrees of freedom per site, the QP band remains visible  throughout the entire BZ and  approaches the lower edge of the continuum near  the edge of the BZ. 

This behavior is in a sharp contrast to  the HM case where the QP band  extends through the whole BZ  and is entirely located  below the continuum. \cite{Bonca_2019,fehske2000,hohen2003,berciu07a,berciu07b,goodvin} Within the incoherent part of the spectra we observe a  rather well defined band of high-energy excitations with its maximal intensity around the edges of the BZ. The separation between between the QP band and these high--energy excitation can be estimated from the energies computed in the  atomic limit of the model, {\it i.e.} at $t_0=0$
\begin{equation}
\epsilon_{t_0=0}^{\pm}={1\over 2}(\omega_0\pm\sqrt{\omega_0^2+4g^2})
\end{equation}
that gives the energy gap $\Delta_\mathrm{gap}=\sqrt{\omega_0^2+4g^2}=4.1$, which further determines an  estimate of the separation between the two bands.  With increasing $T$ additional spectral weight develops below the QP band predominantly around the center of the BZ, as seen in Figs.~\ref{Fig6}(b) and (c). This buildup of the spectra below the QP band  appears because at elevated $T$ an   electron  can annihilate  a thermally excited  HCB. 

Remnants of the QP band remain clearly distinguishable  even at $T=\omega_0$, the same holds also for   the well pronounced  peaks around the edges of the BZ in the high-frequency regime.  In contrast to the HM case, the extra spectral weight that develops below the QP band extends only $\omega_0$ below the bottom of the QP band consistent with the fact that only a single HCB is allowed per site. 

In spectral functions at small $\omega_0=0.2$, shown in Figs.~\ref{Fig6}(d) through (f),   the QP band is barely visible even at small $T/\omega_0=0.05$, it merges with the continuum around $k\sim 0.35\pi$. This is also consistent with the behavior of  $Z_\mathrm{qp}$, shown in Fig.~\ref{fig1}(f), that at $k\sim 0.35\pi$ approaches zero. In contrast to the $\omega_0=1$ case where the QP band is well separated from the incoherent continuum,  at $\omega_0=0.2$ we find rather large spectral weight just above the QP band. Towards the edge of the BZ the incoherent part of the spectrum consists of a series of separated and split  bands.  


\begin{figure}[!tbh]
\includegraphics[width=1.0\columnwidth]{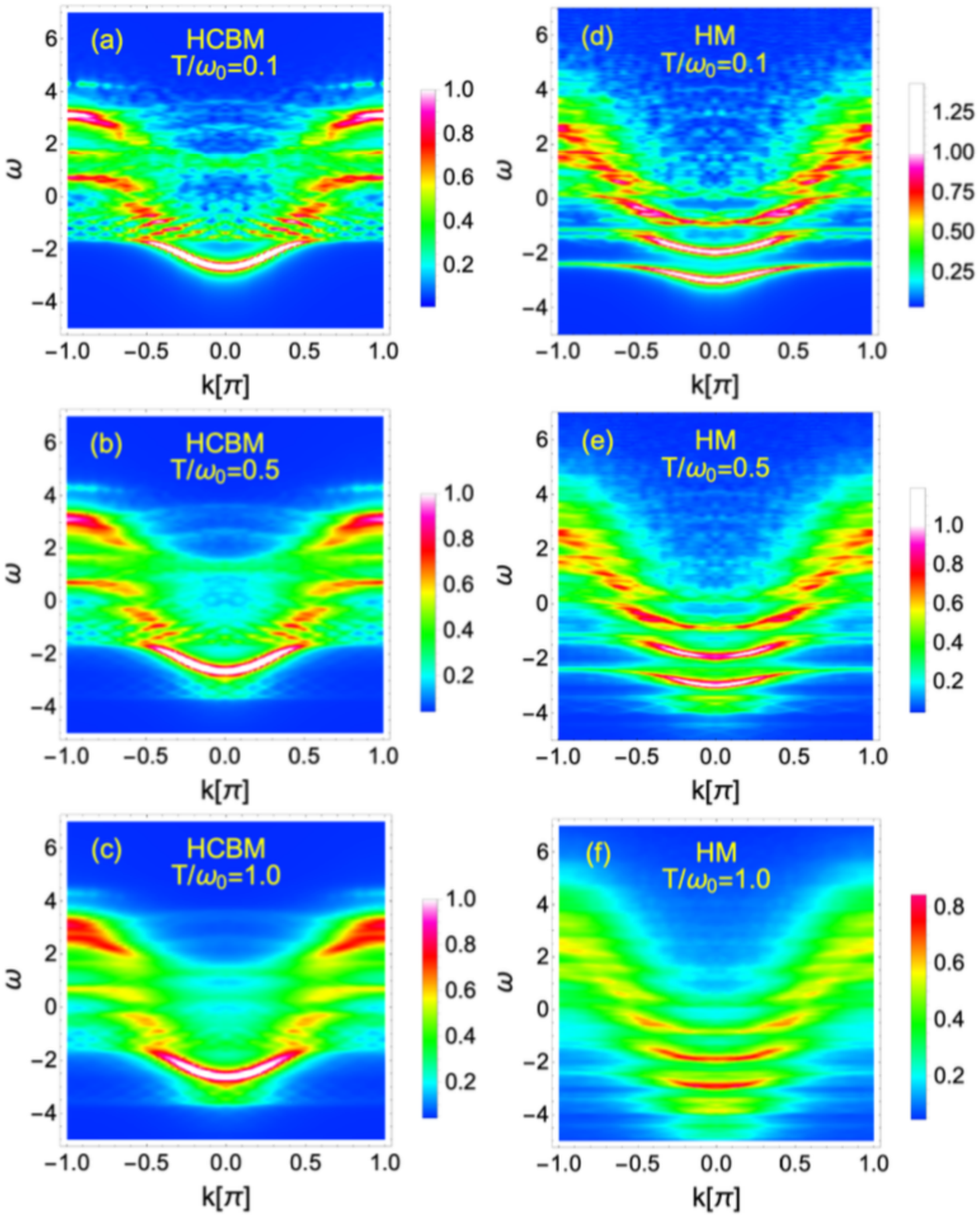}
\caption{Comparison of  $A(\omega,k)$ for $\omega_0=1$ and $g=\sqrt{2}$ or $\lambda=g^2/(2\omega_0t_0)=1$, vs. different values of $T/\omega_0$ fort two different models. The dimensionless coupling constant $\lambda$  in the HM case at $\lambda=1$ represents the intermediate coupling regime. From (a) through (c), for the HCBM and from (d) through (f) for the HM. The latter set of results was computed  using Hamiltonian as defined in Eq.~\ref{ham} but with  standard boson commutation relations $[b_i,b_j^\dagger]=\delta_{i,j}$ and by using the method, described in Ref.~\cite{Bonca_2019}. 
In all figures from (a) to (f) we have used  identical color coding  to enable direct comparison between different cases.  Lorentzian broadening $\eta=0.05$ was used in all cases.   }
\label{Fig7}
\end{figure}


   In Fig.~\ref{Fig7} we show comparison  between  $A(\omega,k)$ as obtained from the HCBM and the standard HM using   identical model parameters. Most notable distinction is observed already at small $T/\omega_0=0.1$, as shown  in Figs.~\ref{Fig7}(a) and (d). While in the HCBM  a single well defined QP band  in the vicinity of  the center of the BZ is observed, in the HM  there are multiple well defined bands separated by $\omega_0$ above the QP band extending throughout  the entire BZ. Spacing between multiple bands observed in the case of HM model can be explained by inspecting   the energy spectra in the atomic limit of the model given by $\epsilon_\mathrm{t_0=0}^n=-g^2/\omega_0 + n \omega_0$.  In addition we find larger  effective mass in the HM that is reflected in a less dispersive QP band in comparison to the HCB case. A quantitative analysis yields  a  mass ratio $m_\mathrm{eff}^\mathrm{HM}/m_\mathrm{eff}^\mathrm{HCB}\sim 1.6$.

Shifting now the comparison to finite-$T$ results we observe additional spectral weight that emerges with increasing $T$ below the respective QP bands in both cases. Nevertheless, in the HCBM the latter is limited to the energy interval $[E(k), E(k)- \omega_0]$ and remains restricted  to the center of the BZ  while in the HM case it appears as replicas of the nearly flat QP bands located at energies $E(k)-n\omega_0$ and with decreasing intensity as  $n\in \cal N$ increases.

\subsection{Self--energies}

\begin{figure}[!tbh]
\includegraphics[width=1.0\columnwidth]{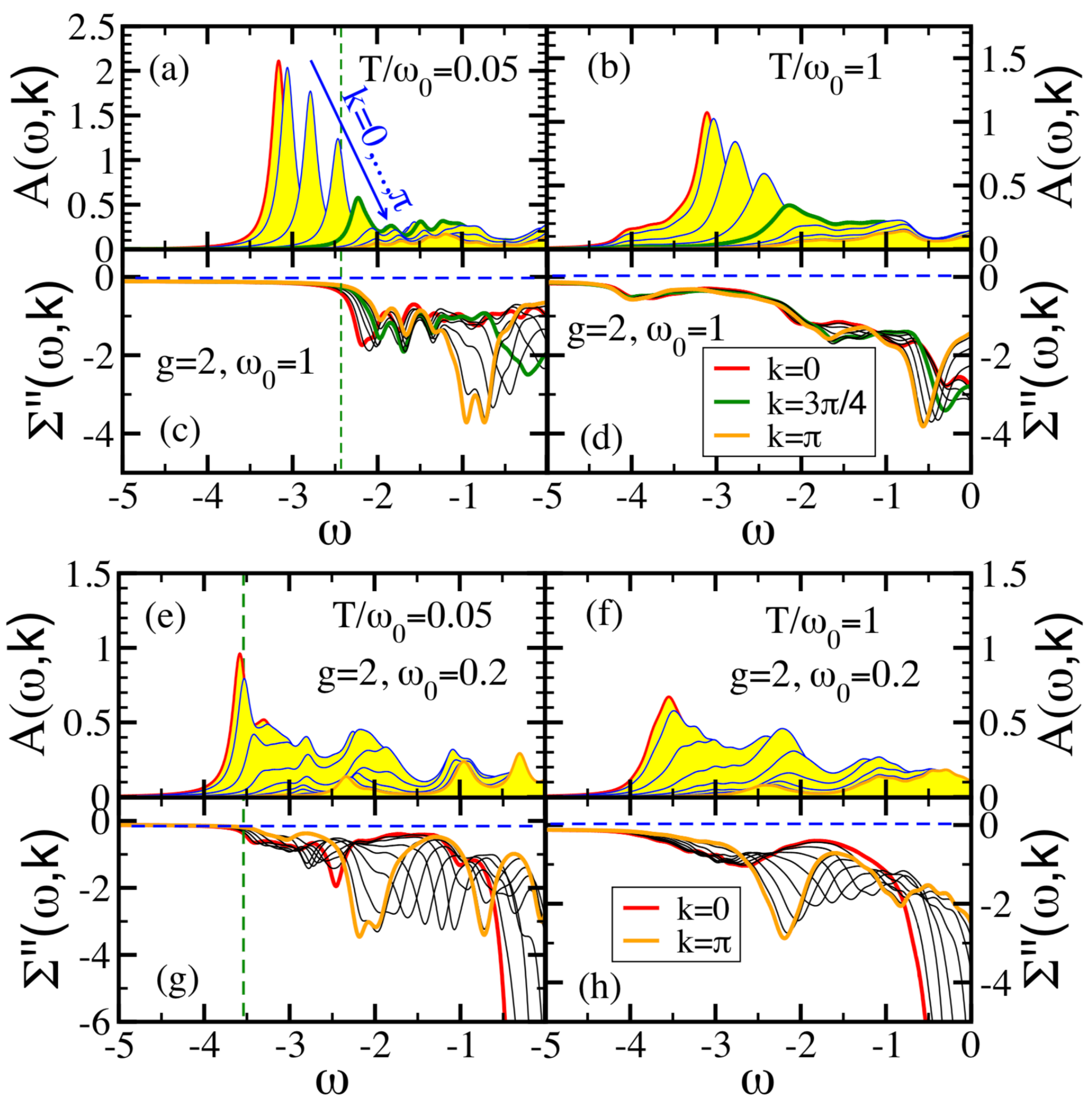}
\caption{$A(\omega,k)$ and $\Sigma(\omega,k)$ for $g=2$ and $\omega_0=1$ and 0.2 computed at two  distinct  values of $T/\omega_0=0.1$ in (a,c,e,g) and 1.0  in (b,d,f,h). 
The same size of the system was used as in Figs.~\ref{fig1} through \ref{Fig6}. In each figure we present  9 curves computed at  $k=m\pi/8$, $m=0,\dots, 8$.  Lorentzian broadening $\eta=0.05$ was used in all cases, which is also responsible for a small deviation from zero in Fig.~\ref{Fig8} (c) for $\omega\lesssim -2.5$. }
\label{Fig8}
\end{figure}

In order to study the life--time of the QP peaks at elevated $T$, we have computed the corresponding self--energies. 
In Fig.~\ref{Fig8} we show a direct comparison between  a family of $A(\omega,k)$ obtained in the interval  $k\in [0,\pi]$ and  the  imaginary parts of matching  self--energies $\Sigma^{''}(\omega,k)$ for two different values of $\omega_0=1$ and $0.2$  at low--$T/\omega_0=0.05$ as well as for  high--$T/\omega_0=1$.  Results of $\Sigma^{''}(\omega,k)$ were obtained   from the relation ${\cal G}^R(\omega,k)=1/(\omega-\epsilon(k)-\Sigma(\omega,k))$ where $\epsilon(k)$ is the free  electron energy $\epsilon(k)=-2t_0\cos(k)$.  
From comparison of Figs.~\ref{Fig8}(a) and (c) we notice that $\Sigma^{''}(\omega,k)\sim 0$ in the frequency regime that corresponds to the well defined QP band in Fig.~\ref{fig1}(a). In Fig.~\ref{Fig8}(a) this regime however does not extend though the entire BZ since $\Sigma^{''}(\omega,k)$  becomes finite already around $k_0=3\pi/4$ and at the  value of $\omega$ corresponding to the lowest peak in  $A(\omega,k_0)$. This behavior shows that a   well defined quasiparticle with an  infinitely long lifetime exists only in a limited region  of the BZ. In Fig.~\ref{Fig8}(g), corresponding to $\omega_0=0.2$ the regime of $\Sigma^{''}(\omega,k)\sim 0$ is even more limited to the vicinity of the QP peak at $k=0$. In both cases  $\Sigma^{''}(\omega,k)$ displays rather pronounced  $k-$ dependence which is in contrast to the HM case.\cite{Bonca_2019} At elevated $T$ $\Sigma^{''}(\omega,k)$ deviates significantly from zero already  at frequencies, corresponding to the positions of QP peaks as seen in Figs.~\ref{Fig8}(d) and (h). This result   demonstrates   the importance of incoherent processes that originate from thermal collisions.

\subsection{Frequency moments}
\label{FM}

We have computed the lowest frequency  moments of the spectral function.  As already  shown  in Ref.~\cite{kornilovitch,goodvin,Bonca_2019}, frequency moments of the single polaron spectral function can be obtained analytically using the following relation
\begin{eqnarray}
M_m(k) &=& \int_{-\infty}^\infty  \omega^mA(\omega,k)\mathrm{d}\omega=\nonumber\\ 
             &=&\langle [[[c_k,H],H],\dots,H]c_k^\dagger\rangle_T,
\end{eqnarray}
where $\langle \dots\rangle_T$ represents the thermal average over zero--electron states and the number of commutators corresponds to the order of the frequency moment. Taking into account the HCB commutation relation $[b_i,b^\dagger_j]=\delta_{i,j}(1-2b^\dagger_ib_i)$ analytical expressions may be  obtained for arbitrary moments even  at finite-$T$. Here we list just a few:
\begin{eqnarray}
M_0(k)&=&1,\nonumber\\
M_1(k)&=&\epsilon(k),\nonumber\\
M_2(k)&=&\epsilon^2(k) + g^2,\nonumber\\
M_3(k)&=&\epsilon^3(k)+2g^2\epsilon(k) + g^2\omega_0(1-2n_b^0),\label{mom}
\end{eqnarray}
where $\epsilon(k)=-2t_0\cos(k)$ and $n_b^0=1/(\exp(\omega_0/T)+1)$. Note that $M_0(k)$ through  $M_2(k)$ do not depend on $T$.

\begin{figure}[!tbh]
\includegraphics[width=0.8\columnwidth]{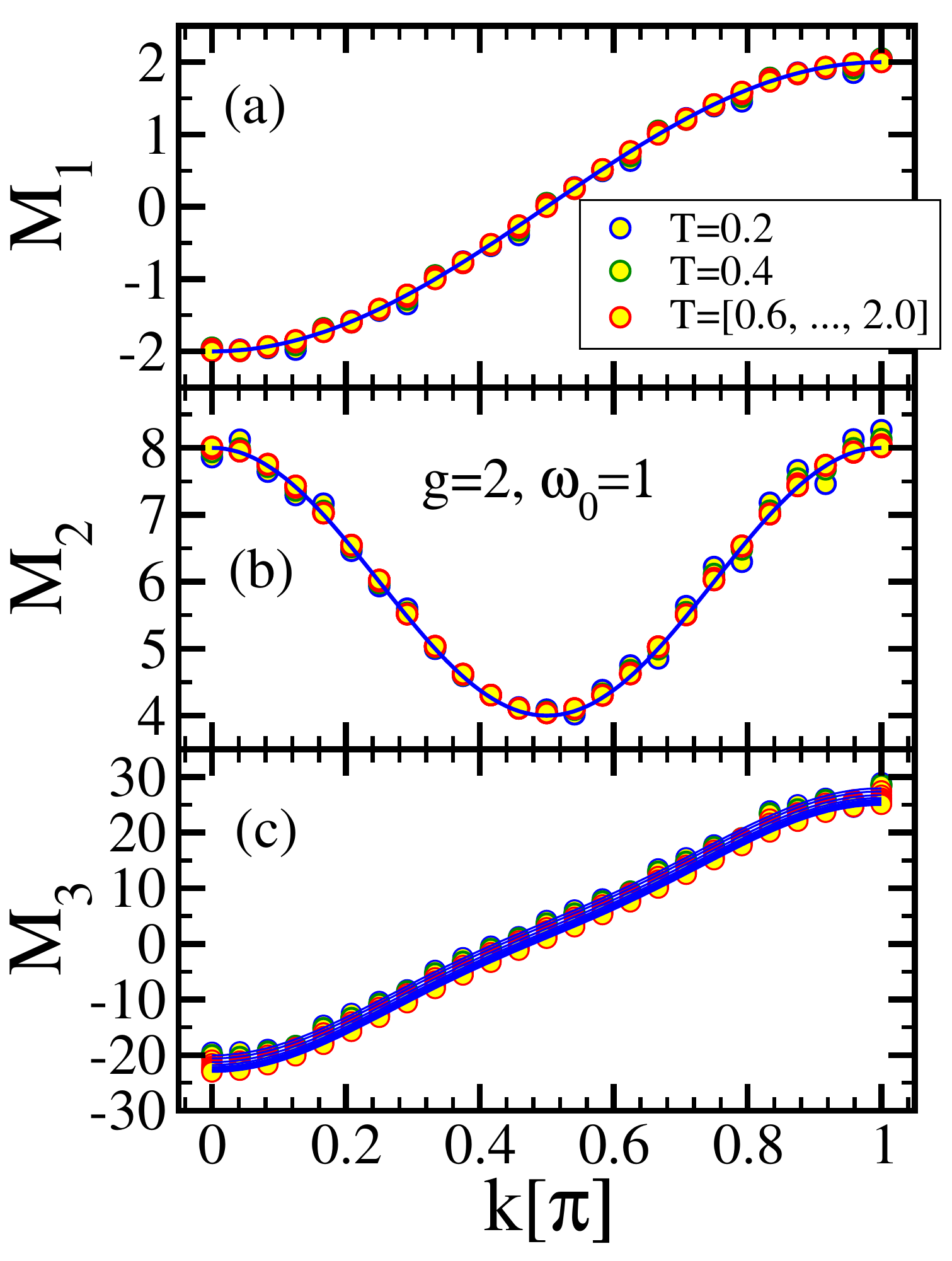}
\caption{$M_1(k)$, $M_2(k)$ and $M_3(k)$ at $\omega_0=1$ and $g=2$.  In all plots we present with  open symbols frequency   moments obtained from  numerical results of $A(\omega,k)$ computed on $L=16$ sites system with twisted boundary conditions, just as in Fig.~\ref{Fig6},    but for  10 different values of $T\in [0.2, 0.4,\dots, 2.0]$. Results for  the lowest $T=0.2$ and 0.4 are represented with blue and green circles, respectively, while the rest of results for $T\geq 0.6$  are shown  in red. Blue lines represent analytic results from Eq.~\ref{mom}.  In (c)  we observe a slight downward shift of numerical data with increasing  $T$, which is   nearly perfectly captured  by  a set of thin lines representing $M_3(k)$,   Eq.~\ref{mom}.  \label{Fig9}}
\end{figure}

In Figs.~\ref{Fig9} we plot $M_n(k);~n\in[1,2,3]$  extracted from $A(\omega,k)$ for a set of temperatures $T\in[0.2, 0.4, \dots 2.0]$. Except for  the smallest $T=0.2$ and 0.4, numerically obtained frequency moments match nearly perfectly analytical predictions from Eq.~\ref{mom} that are free of finite-size effects.  A slight disagreement  at low-$T$  could be the consequence of the discrete as well as limited   frequency interval used in our calculations. It is instructive to note, that the agreement with analytical predictions becomes even better at higher-$T$ which further validates the applicability of the Finite-$T$ Lanczos procedure as well as the introduction of twisted boundary conditions. The attention should also be drawn to the notion that  in contrast to numerically obtained $M_n(k)$ from calculations on a finite size system with $L=16$, analytical results in Eq.~\ref{mom} are free of finite--size effects. By applying twisted boundary conditions we were able to present $M_n(k)$ on 25 $k$--points spaced by $\Delta k = \pi/24$ instead of on  $L/2+1=9$ points when using periodic boundary conditions. We observe no spurious effects that might be expected due to twisted boundary conditions.

\section{Summary}

We have computed  typical ground state as well as finite-$T$ static and dynamic properties of an electron coupled to inelastic HCB degrees of freedom. While the model resembles a widely  investigated HM, the limitation of infinite  phonon degrees of freedom to  HCB's brings about important differences. A polaron, dressed with HCB excitations remains light, its effective mass remains of the order of the free electron mass even in the limit of strong coupling. Still, the model remains within  a class of non-integrable models that exhibit ergodic properties at elevated temperatures. 

In the electron spectral function we observe a dispersive  QP peak forming a QP band with its maximal spectral   weight centered around the center of the BZ. The peak  remains separated from the continuum of states only in the central part of the BZ. With increasing momentum the weight of the QP band approaches zero as the  peak enters  the continuum while   the imaginary part of the self--energy starts deviating from zero, which  in turn signals the appearance of a finite lifetime. This result is in contrast to the HM where the QP band  remains separated from the continuum throughout the  whole BZ.  With increasing $T$ additional features  in the spectral function emerge  in a form of an additional spectral weight below the QP band   already at temperatures below $\omega_0$. These features  appear   due to  processes where an electron annihilates one or more   thermally excited HCB's. 

As mentioned in the introduction, models where electrons are coupled to HCB's may carry potential relevance to microscopic mechanisms of superconductivity where the attraction between electrons is mediated by HCB degrees of freedom. Such mechanisms have been comprehensively  investigated in connection to bipolaron formation mediated by lattice vibrations.\cite{alex_book_1995,alexandrov,alexandrov07,alexandrov94,alexandrov95,bonca2000}  One of the shortcomings of  these theories  are exponentially large effective polaron and bipolaron masses in the strong  electron phonon coupling regime.  We have also shown that in a system of two electrons with  total spin $S=0$ and in the absence of the on-site Coulomb interaction a bound bipolaron exists at any finite HCB--electron coupling strength $g$. In the case of finite Coulomb interaction there exists a threshold value of $g_c$.  Since the  effective polaron mass remains small even in the limit of very strong electron-HCB coupling regime, HCB--mediated electron electron  interaction  leads to formation of light bipolarons.

\begin{acknowledgments}
I  acknowledge the support by the program P1-0044 of the
Slovenian Research Agency, support from  the Centre for Integrated Nanotechnologies, a U.S. Department of Energy, Office of Basic Energy Sciences user facility, and funding from the Stewart Blusson Quantum Matter Institute. I also acknowledge the hospitality of dr. M. Stout and  loving support of my wife J. Bon\v ca Vidmar. 
\end{acknowledgments}

\setcounter{figure}{0}
\setcounter{table}{0}
\makeatletter
\renewcommand{\theequation}{A\arabic{equation}}
\renewcommand{\thefigure}{A\arabic{figure}}
 
 \appendix
 
\section{ }
\label{A}
\begin{figure}[!tbh]
\includegraphics[width=0.8\columnwidth]{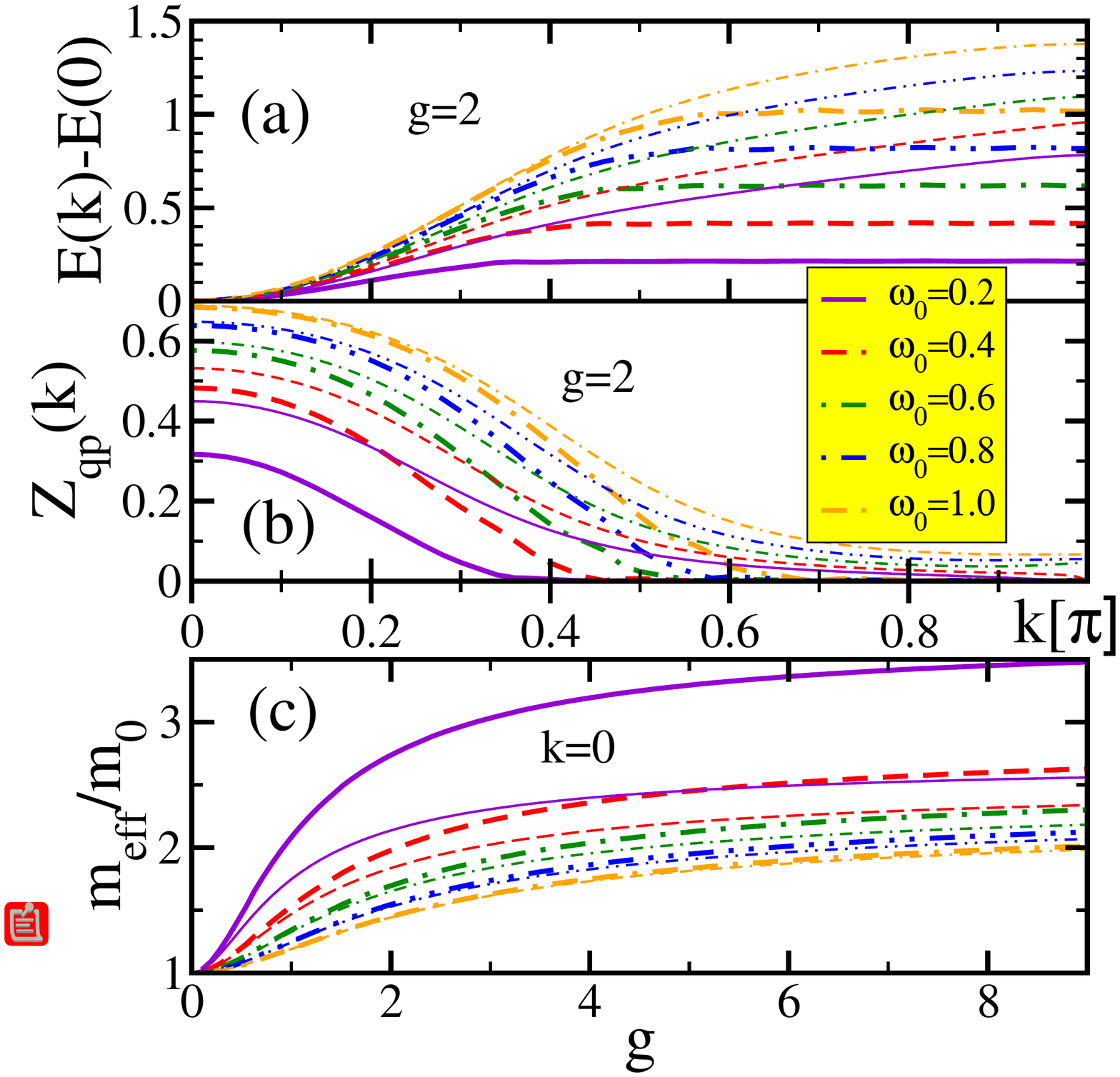}
\caption{ Ground-state properties computed using full basis on a ring with $L=16$ with $N_\mathrm{st}=2^L$ parent states (thick lines)  taking into account the  full translational symmetry  in comparison with calculations using only $N_\mathrm{st}=14$ parent states (thin lines):  a) and b):  $k-$ dependent energy $E(k)-E(0)$ and $Z_\mathrm{qp}(k)$, respectively,  at fixed coupling $g$;  c) the effective mass $m_\mathrm{eff}/m_0=2t_0(\partial ^2 E(k)/\partial k^2\vert_{k=0}  )^{-1}$ vs. $g$.  }
\label{figS1}
\end{figure}

To gain further physical intuition into the ground--state properties of the HCBM as well as into the numerical approach used in this work, we have diagonalised the model in a  restricted translationally invariant basis  that spans only 14 states,  obtained   by acting with the off-diagonal part of Hamiltonian in Eq.~\ref{ham}, {\it i.e.}
\begin{eqnarray}
H_\mathrm{off} &=&H_\mathrm{t_0}+H_g=\nonumber \\
& -&t_0\sum_{j}(c^\dagger_{j} c_{j+1} +\mathrm{H.c.})-{g} \sum_{j} \hat n_{j} (b_{j}^\dagger + b_{j}),   \label{off}
\end{eqnarray}
$N_h-$ times on a state with a single electron in a translationally invariant state $k$ with  zero HCB degrees of freedom to obtain a limited basis set
\begin{equation}
\left \{ \vert \phi_{{ k},l}^{(N_h)}{\rangle}\right \} =H_\mathrm{off}^{N_h} c_{ k}^{\dagger} \vert \emptyset{\rangle}. 
\label{gen}
\end{equation}
Translationally invariant basis states are represented by HCB position coordinates $i_j$ as $\vert i_1,i_2,\dots,i_{N_\mathrm{HCB}}\rangle_k,$ where $N_\mathrm{HCB}$ denotes  the number of HCB's of  a particular parent state. Due to the translational invariance, the electron position is kept fixed and does not need to be indexed.  For the case of $N_h=4$ we obtain 14 states: $\vert \emptyset{\rangle}_k=c_k^\dagger \vert\emptyset\rangle, \vert 0\rangle_k, \vert \pm1\rangle_k,\vert 0,\pm 1\rangle_k,\vert \pm 2\rangle_k,\vert \pm 3\rangle_k, \vert 0,\pm 2\rangle_k,\vert1,2\rangle_k,\vert -1,-2\rangle_k$. We next list just a few non--zero matrix elements:
\begin{eqnarray}
_k\langle \emptyset \vert H \vert\emptyset\rangle_k&=&-2t_0\cos(k);  \\
{_k\langle} 0\vert H \vert 0\rangle_k&=&{_k\langle}\pm1\vert H\vert \pm1\rangle_k=\omega_0\dots\nonumber\\
_k\langle \emptyset \vert H \vert 0\rangle_k&=&{_k\langle} \pm 1 \vert H \vert 0\pm 1\rangle_k=-g;\nonumber \\
{_k\langle} 0\vert H\vert \pm 1\rangle_k &=& {_k\langle} \pm 1\vert H\vert \pm 2\rangle_k=-t_0e^{\mp ik} \nonumber \\
_k\langle 0,\pm1\vert H\vert 0,\pm 1\rangle_k&=&2\omega_0.\nonumber\\
\dots \nonumber
\end{eqnarray}
In Fig.~\ref{figS1} we show a comparison between results obtained using a full translationally invariant basis (FB) on a $L=16$ sites system with $N_\mathrm{st}=2^L$ states with those obtained using only 14 limited basis states (LBS) as listed above.  Note also that results presented in the main body of the paper have been obtained using the full translationally invariant  basis. The only  exemption from this rule are results of the HM presented in Fig.~\ref{Fig7} where LBS states have been used combined with the standard Lanczos technique\cite{lanczos} and described in details in Refs.~\cite{bonca1,bonca4}.  Using LBS we expectedly obtain consistently higher energies in comparison to the FB while it is also  evident that differences between results decrease with increasing $\omega_0$ at fixed coupling $g$.   This consistently holds true for all quantities, shown in Fig.~\ref{figS1}. It is rather surprising that the comparison of  $m_\mathrm{eff}/m_0$ for the largest  $\omega_0=1$  shows nearly identical results obtained using  substantially  different numbers of basis states. Some of the main characteristics of ground--state properties may be discerned even from considerably  reduced LBS, such as a decrease of $Z_\mathrm{qp}(k)$ with increasing $k$ and a slow increase of $m_\mathrm{eff}/m_0$ with $g$. Nevertheless,  much larger systems with complete basis states are needed to obtain some of the most interesting properties of the polaron, such as the disappearance of the $Z_\mathrm{qp}(k)$ with increasing $k$, and crossing of the QP band with the continuum of states that is reflected in the flattening of $E(k)$ at $k_0$, given by the solution of $E(k_0)=E(0)+\omega_0$. In addition, using FB as opposed to LBS is necessary to investigate finite--$T$ properties of the model where multiple HCB excitations are needed to properly describe thermally activated processes. 

\setcounter{figure}{0}
\setcounter{table}{0}

\renewcommand{\theequation}{B\arabic{equation}}
\renewcommand{\thefigure}{B\arabic{figure}}
\section{Finite-size analysis}
\label{B}
 
\begin{figure}[!tbh]
\includegraphics[width=1.0\columnwidth]{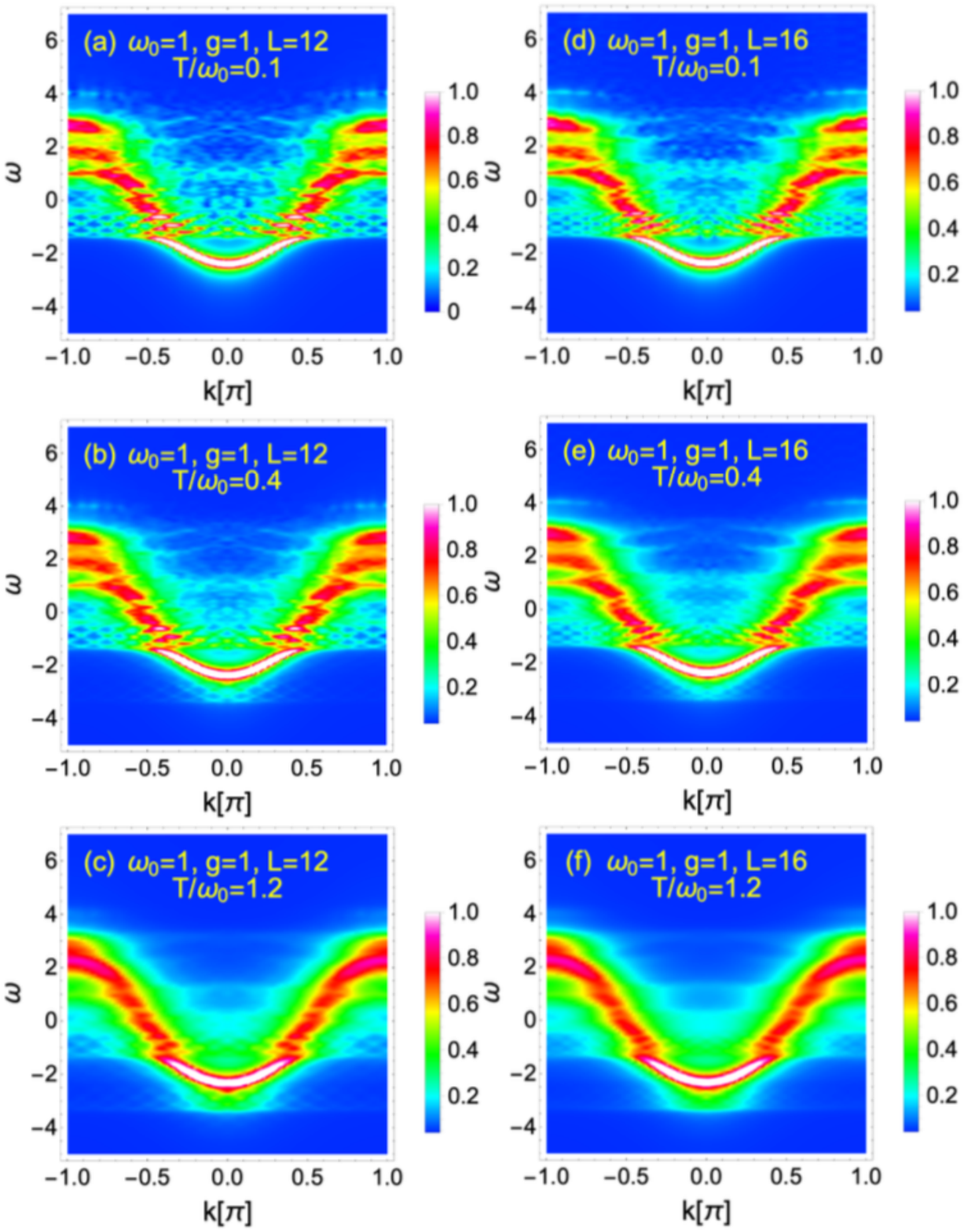}
\caption{ $A(\omega,k)$ for $\omega_0=1, g=2$  for two different system sizes $L=12$ shown in (a) through  (c)  and $L=16$ shown in (d), through (f) and three  different values of $T/\omega_0$ as specified in figures.   In all cases   $A(\omega,k)$  was computed in  25 equally spaced non--equivalent $k-$ points in the interval $k\in [0,\pi]$ with increments of $\Delta k = \pi/24$. Note, in all figures from (a) to (f) the same color coding was used to enable direct comparison between different cases.
  }
\label{figS2}
\end{figure}
We investigate the extend of finite--size effects on the spectral function. In Fig.~\ref{figS2} we present results obtained on two different systems with $L=12$ and 16 sites. In both cases $A(\omega,k)$ were computed on discrete $k-$ points according to periodic as well as twisted  boundary conditions,  equivalent to $k_{n,m} = 2\pi n/L  + m \theta$ with   $n\in [-L/2,L/2]$ and   $\theta = 2\pi/(M_\theta L)$; $m\in [0,M_\theta -1]$. For $L=12$ and 16 systems we have chosen $M=4$ and 3, respectively.  As a result, for each system size  $A(\omega,k)$ were computed using $M_\theta *L/2+1 $ nonequivalent k- points.  Despite substantially different system sizes results are qualitatively identical at any $T$.

\begin{figure}[!tbh]
\includegraphics[width=1.0\columnwidth]{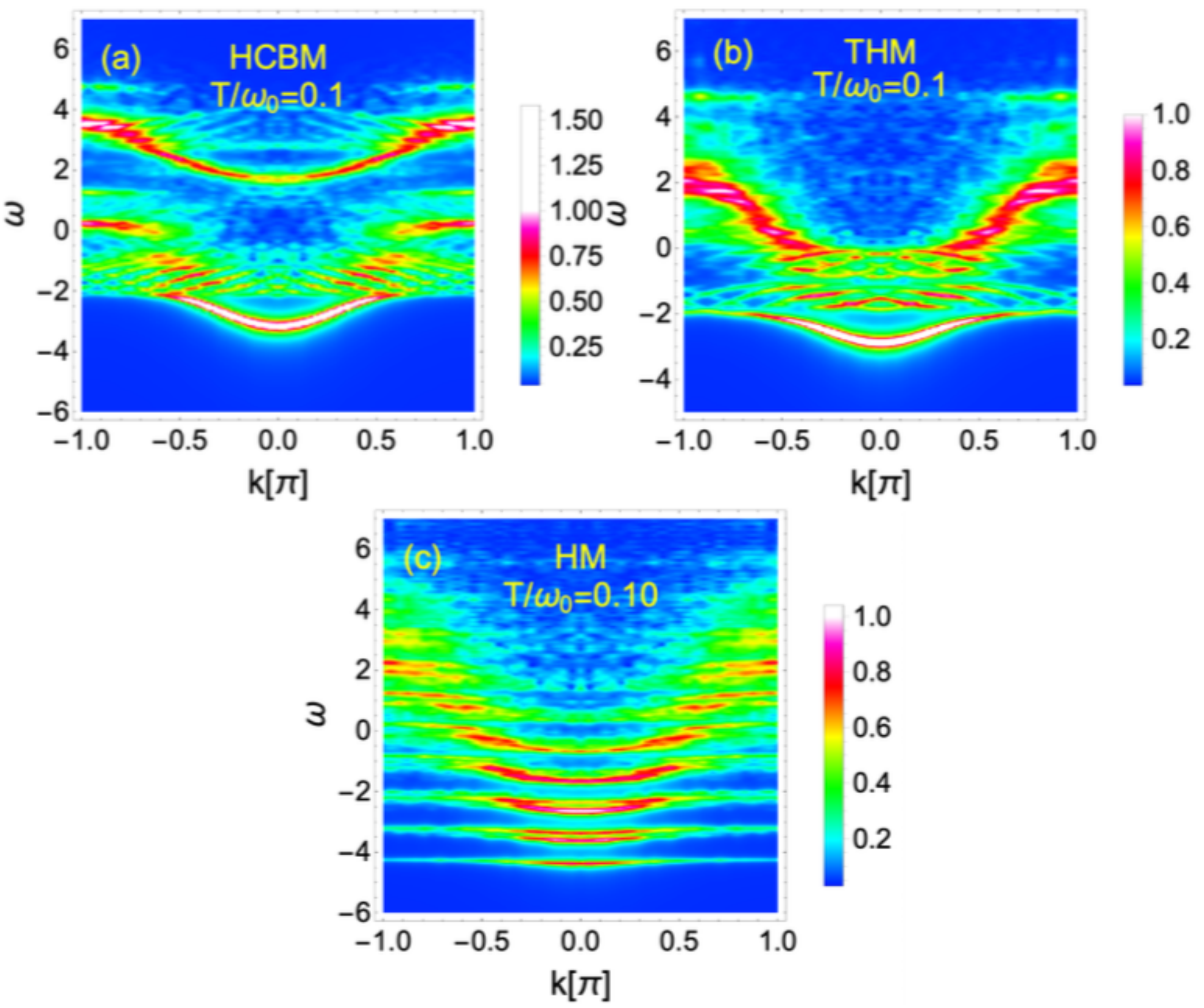}
\caption{ $A(\omega,k)$ for $\omega_0=1, g=2$  for a) the HCBM, b) the truncated HM (THM) with up to  2 phonon quanta per site , and c) the HM with up to 22 phonon quanta per site.  
  }
\label{figS3}
\end{figure}

Last, we follow the evolution of $A(\omega,k)$ with increasing number of allowed bosonic excitations per site starting from the HCBM  in Fig.~\ref{figS3}(a), over  a truncated  HM (THM) where we allow up to 2 phonon quanta per site,  to the HM in Fig.~\ref{figS3}(c) where we have expanded the Hilbert space up to maximal 22 phonon excitations per site. The emphasis of this analysis is on the evolution of the QP  band. While in the HCBM the QP band  enters the continuum at finite $k_0$, already in the HMR  it flattens out and  shows a tendency to disperse  below the continuum almost towards  the edge of the BZ, as seen in Fig.~\ref{figS3}(b). Note that in the case of the HM $g=2$ represents the strong coupling limit, $\lambda=2$, where we observe a nearly flat QP band with a substantially reduced QP weight,  followed by a series of nearly flat bands, separated  by $\omega_0$ as predicted by the strong coupling theory. 

\bibliography{manuspe}

\begin{thebibliography}{40}
\expandafter\ifx\csname natexlab\endcsname\relax\def\natexlab#1{#1}\fi
\expandafter\ifx\csname bibnamefont\endcsname\relax
  \def\bibnamefont#1{#1}\fi
\expandafter\ifx\csname bibfnamefont\endcsname\relax
  \def\bibfnamefont#1{#1}\fi
\expandafter\ifx\csname citenamefont\endcsname\relax
  \def\citenamefont#1{#1}\fi
\expandafter\ifx\csname url\endcsname\relax
  \def\url#1{\texttt{#1}}\fi
\expandafter\ifx\csname urlprefix\endcsname\relax\def\urlprefix{URL }\fi
\providecommand{\bibinfo}[2]{#2}
\providecommand{\eprint}[2][]{\url{#2}}

\bibitem[{\citenamefont{Holstein}(1959)}]{holstein59}
\bibinfo{author}{\bibfnamefont{T.}~\bibnamefont{Holstein}},
  \bibinfo{journal}{Annals of Physics} \textbf{\bibinfo{volume}{8}},
  \bibinfo{pages}{325 } (\bibinfo{year}{1959}), ISSN \bibinfo{issn}{0003-4916},
  \urlprefix\url{http://www.sciencedirect.com/science/article/pii/0003491659900028}.

\bibitem[{\citenamefont{Engelsberg and Schrieffer}(1963)}]{engelsberg63}
\bibinfo{author}{\bibfnamefont{S.}~\bibnamefont{Engelsberg}} \bibnamefont{and}
  \bibinfo{author}{\bibfnamefont{J.~R.} \bibnamefont{Schrieffer}},
  \bibinfo{journal}{Phys. Rev.} \textbf{\bibinfo{volume}{131}},
  \bibinfo{pages}{993} (\bibinfo{year}{1963}),
  \urlprefix\url{https://link.aps.org/doi/10.1103/PhysRev.131.993}.

\bibitem[{\citenamefont{Bon\v{c}a et~al.}(1999)\citenamefont{Bon\v{c}a,
  Trugman, and Batisti\'{c}}}]{bonca1}
\bibinfo{author}{\bibfnamefont{J.}~\bibnamefont{Bon\v{c}a}},
  \bibinfo{author}{\bibfnamefont{S.~A.} \bibnamefont{Trugman}},
  \bibnamefont{and}
  \bibinfo{author}{\bibfnamefont{I.}~\bibnamefont{Batisti\'{c}}},
  \bibinfo{journal}{Phys. Rev. B} \textbf{\bibinfo{volume}{60}},
  \bibinfo{pages}{1633} (\bibinfo{year}{1999}).

\bibitem[{\citenamefont{Ku et~al.}(2002)\citenamefont{Ku, Trugman, and
  Bon\v{c}a}}]{bonca4}
\bibinfo{author}{\bibfnamefont{L.~C.} \bibnamefont{Ku}},
  \bibinfo{author}{\bibfnamefont{S.~A.} \bibnamefont{Trugman}},
  \bibnamefont{and}
  \bibinfo{author}{\bibfnamefont{J.}~\bibnamefont{Bon\v{c}a}},
  \bibinfo{journal}{Phys. Rev. B} \textbf{\bibinfo{volume}{65}},
  \bibinfo{pages}{174306} (\bibinfo{year}{2002}).

\bibitem[{\citenamefont{Ranninger and Thibblin}(1992)}]{Ranninger1992}
\bibinfo{author}{\bibfnamefont{J.}~\bibnamefont{Ranninger}} \bibnamefont{and}
  \bibinfo{author}{\bibfnamefont{U.}~\bibnamefont{Thibblin}},
  \bibinfo{journal}{Phys. Rev. B} \textbf{\bibinfo{volume}{45}},
  \bibinfo{pages}{7730} (\bibinfo{year}{1992}),
  \urlprefix\url{https://link.aps.org/doi/10.1103/PhysRevB.45.7730}.

\bibitem[{\citenamefont{Marsiglio}(1993)}]{marsiglio93}
\bibinfo{author}{\bibfnamefont{F.}~\bibnamefont{Marsiglio}},
  \bibinfo{journal}{Physics Letters A} \textbf{\bibinfo{volume}{180}},
  \bibinfo{pages}{280 } (\bibinfo{year}{1993}).

\bibitem[{\citenamefont{Alexandrov et~al.}(1994)\citenamefont{Alexandrov,
  Kabanov, and Ray}}]{alexandrov94}
\bibinfo{author}{\bibfnamefont{A.~S.} \bibnamefont{Alexandrov}},
  \bibinfo{author}{\bibfnamefont{V.~V.} \bibnamefont{Kabanov}},
  \bibnamefont{and} \bibinfo{author}{\bibfnamefont{D.~K.} \bibnamefont{Ray}},
  \bibinfo{journal}{Phys. Rev. B} \textbf{\bibinfo{volume}{49}},
  \bibinfo{pages}{9915} (\bibinfo{year}{1994}).

\bibitem[{\citenamefont{Fehske et~al.}(1997)\citenamefont{Fehske, Loos, and
  Wellein}}]{fehske1997}
\bibinfo{author}{\bibfnamefont{H.}~\bibnamefont{Fehske}},
  \bibinfo{author}{\bibfnamefont{J.}~\bibnamefont{Loos}}, \bibnamefont{and}
  \bibinfo{author}{\bibfnamefont{G.}~\bibnamefont{Wellein}},
  \bibinfo{journal}{Zeitschrift f{\"u}r Physik B Condensed Matter}
  \textbf{\bibinfo{volume}{104}}, \bibinfo{pages}{619} (\bibinfo{year}{1997}),
  ISSN \bibinfo{issn}{1431-584X},
  \urlprefix\url{https://doi.org/10.1007/s002570050498}.

\bibitem[{\citenamefont{Fehske et~al.}(2000)\citenamefont{Fehske, Loos, and
  Wellein}}]{fehske2000}
\bibinfo{author}{\bibfnamefont{H.}~\bibnamefont{Fehske}},
  \bibinfo{author}{\bibfnamefont{J.}~\bibnamefont{Loos}}, \bibnamefont{and}
  \bibinfo{author}{\bibfnamefont{G.}~\bibnamefont{Wellein}},
  \bibinfo{journal}{Phys. Rev. B} \textbf{\bibinfo{volume}{61}},
  \bibinfo{pages}{8016} (\bibinfo{year}{2000}),
  \urlprefix\url{https://link.aps.org/doi/10.1103/PhysRevB.61.8016}.

\bibitem[{\citenamefont{Bari\ifmmode \check{s}\else
  \v{s}\fi{}i\ifmmode~\acute{c}\else \'{c}\fi{}}(2002)}]{osor2002}
\bibinfo{author}{\bibfnamefont{O.~S.} \bibnamefont{Bari\ifmmode \check{s}\else
  \v{s}\fi{}i\ifmmode~\acute{c}\else \'{c}\fi{}}}, \bibinfo{journal}{Phys. Rev.
  B} \textbf{\bibinfo{volume}{65}}, \bibinfo{pages}{144301}
  (\bibinfo{year}{2002}),
  \urlprefix\url{https://link.aps.org/doi/10.1103/PhysRevB.65.144301}.

\bibitem[{\citenamefont{Bari\ifmmode \check{s}\else
  \v{s}\fi{}i\ifmmode~\acute{c}\else \'{c}\fi{}}(2004)}]{osor2004}
\bibinfo{author}{\bibfnamefont{O.~S.} \bibnamefont{Bari\ifmmode \check{s}\else
  \v{s}\fi{}i\ifmmode~\acute{c}\else \'{c}\fi{}}}, \bibinfo{journal}{Phys. Rev.
  B} \textbf{\bibinfo{volume}{69}}, \bibinfo{pages}{064302}
  (\bibinfo{year}{2004}),
  \urlprefix\url{https://link.aps.org/doi/10.1103/PhysRevB.69.064302}.

\bibitem[{\citenamefont{Bari\ifmmode \check{s}\else
  \v{s}\fi{}i\ifmmode~\acute{c}\else \'{c}\fi{}}(2006)}]{osor2006}
\bibinfo{author}{\bibfnamefont{O.~S.} \bibnamefont{Bari\ifmmode \check{s}\else
  \v{s}\fi{}i\ifmmode~\acute{c}\else \'{c}\fi{}}}, \bibinfo{journal}{Phys. Rev.
  B} \textbf{\bibinfo{volume}{73}}, \bibinfo{pages}{214304}
  (\bibinfo{year}{2006}),
  \urlprefix\url{https://link.aps.org/doi/10.1103/PhysRevB.73.214304}.

\bibitem[{\citenamefont{De~Filippis et~al.}(2005)\citenamefont{De~Filippis,
  Cataudella, Ramaglia, and Perroni}}]{filippis2005}
\bibinfo{author}{\bibfnamefont{G.}~\bibnamefont{De~Filippis}},
  \bibinfo{author}{\bibfnamefont{V.}~\bibnamefont{Cataudella}},
  \bibinfo{author}{\bibfnamefont{V.~M.} \bibnamefont{Ramaglia}},
  \bibnamefont{and} \bibinfo{author}{\bibfnamefont{C.~A.}
  \bibnamefont{Perroni}}, \bibinfo{journal}{Phys. Rev. B}
  \textbf{\bibinfo{volume}{72}}, \bibinfo{pages}{014307}
  (\bibinfo{year}{2005}),
  \urlprefix\url{https://link.aps.org/doi/10.1103/PhysRevB.72.014307}.

\bibitem[{\citenamefont{Prokof'ev and Svistunov}(1998)}]{prokofev98}
\bibinfo{author}{\bibfnamefont{N.~V.} \bibnamefont{Prokof'ev}}
  \bibnamefont{and} \bibinfo{author}{\bibfnamefont{B.~V.}
  \bibnamefont{Svistunov}}, \bibinfo{journal}{Phys. Rev. Lett.}
  \textbf{\bibinfo{volume}{81}}, \bibinfo{pages}{2514} (\bibinfo{year}{1998}),
  \urlprefix\url{https://link.aps.org/doi/10.1103/PhysRevLett.81.2514}.

\bibitem[{\citenamefont{Cataudella et~al.}(2007)\citenamefont{Cataudella,
  Filippis, Mishchenko, and Nagaosa}}]{mishchenkoSELF}
\bibinfo{author}{\bibfnamefont{V.}~\bibnamefont{Cataudella}},
  \bibinfo{author}{\bibfnamefont{G.~D.} \bibnamefont{Filippis}},
  \bibinfo{author}{\bibfnamefont{A.~S.} \bibnamefont{Mishchenko}},
  \bibnamefont{and} \bibinfo{author}{\bibfnamefont{N.}~\bibnamefont{Nagaosa}},
  \bibinfo{journal}{Phys. Rev. Lett.} \textbf{\bibinfo{volume}{99}},
  \bibinfo{eid}{226402} (pages~\bibinfo{numpages}{4}) (\bibinfo{year}{2007}).

\bibitem[{\citenamefont{Lau et~al.}(2007)\citenamefont{Lau, Berciu, and
  Sawatzky}}]{berciu07a}
\bibinfo{author}{\bibfnamefont{B.}~\bibnamefont{Lau}},
  \bibinfo{author}{\bibfnamefont{M.}~\bibnamefont{Berciu}}, \bibnamefont{and}
  \bibinfo{author}{\bibfnamefont{G.~A.} \bibnamefont{Sawatzky}},
  \bibinfo{journal}{Phys. Rev. B} \textbf{\bibinfo{volume}{76}},
  \bibinfo{pages}{174305} (\bibinfo{year}{2007}).

\bibitem[{\citenamefont{Hohenadler et~al.}(2003)\citenamefont{Hohenadler,
  Aichhorn, and von~der Linden}}]{hohen2003}
\bibinfo{author}{\bibfnamefont{M.}~\bibnamefont{Hohenadler}},
  \bibinfo{author}{\bibfnamefont{M.}~\bibnamefont{Aichhorn}}, \bibnamefont{and}
  \bibinfo{author}{\bibfnamefont{W.}~\bibnamefont{von~der Linden}},
  \bibinfo{journal}{Phys. Rev. B} \textbf{\bibinfo{volume}{68}},
  \bibinfo{pages}{184304} (\bibinfo{year}{2003}),
  \urlprefix\url{https://link.aps.org/doi/10.1103/PhysRevB.68.184304}.

\bibitem[{\citenamefont{Berciu and Goodvin}(2007)}]{berciu07b}
\bibinfo{author}{\bibfnamefont{M.}~\bibnamefont{Berciu}} \bibnamefont{and}
  \bibinfo{author}{\bibfnamefont{G.~L.} \bibnamefont{Goodvin}},
  \bibinfo{journal}{Phys. Rev. B} \textbf{\bibinfo{volume}{76}},
  \bibinfo{pages}{165109} (\bibinfo{year}{2007}).

\bibitem[{\citenamefont{Goodvin et~al.}(2006)\citenamefont{Goodvin, Berciu, and
  Sawatzky}}]{goodvin}
\bibinfo{author}{\bibfnamefont{G.~L.} \bibnamefont{Goodvin}},
  \bibinfo{author}{\bibfnamefont{M.}~\bibnamefont{Berciu}}, \bibnamefont{and}
  \bibinfo{author}{\bibfnamefont{G.~A.} \bibnamefont{Sawatzky}},
  \bibinfo{journal}{Phys. Rev. B} \textbf{\bibinfo{volume}{74}},
  \bibinfo{pages}{245104} (\bibinfo{year}{2006}).

\bibitem[{\citenamefont{Bon\ifmmode~\check{c}\else \v{c}\fi{}a
  et~al.}(2019)\citenamefont{Bon\ifmmode~\check{c}\else \v{c}\fi{}a, Trugman,
  and Berciu}}]{Bonca_2019}
\bibinfo{author}{\bibfnamefont{J.}~\bibnamefont{Bon\ifmmode~\check{c}\else
  \v{c}\fi{}a}}, \bibinfo{author}{\bibfnamefont{S.~A.} \bibnamefont{Trugman}},
  \bibnamefont{and} \bibinfo{author}{\bibfnamefont{M.}~\bibnamefont{Berciu}},
  \bibinfo{journal}{Phys. Rev. B} \textbf{\bibinfo{volume}{100}},
  \bibinfo{pages}{094307} (\bibinfo{year}{2019}),
  \urlprefix\url{https://link.aps.org/doi/10.1103/PhysRevB.100.094307}.

\bibitem[{\citenamefont{Jansen et~al.}(2019)\citenamefont{Jansen, Stolpp,
  Vidmar, and Heidrich-Meisner}}]{Vidmar_2019}
\bibinfo{author}{\bibfnamefont{D.}~\bibnamefont{Jansen}},
  \bibinfo{author}{\bibfnamefont{J.}~\bibnamefont{Stolpp}},
  \bibinfo{author}{\bibfnamefont{L.}~\bibnamefont{Vidmar}}, \bibnamefont{and}
  \bibinfo{author}{\bibfnamefont{F.}~\bibnamefont{Heidrich-Meisner}},
  \bibinfo{journal}{Phys. Rev. B} \textbf{\bibinfo{volume}{99}},
  \bibinfo{pages}{155130} (\bibinfo{year}{2019}),
  \urlprefix\url{https://link.aps.org/doi/10.1103/PhysRevB.99.155130}.

\bibitem[{\citenamefont{Brinkman and Rice}(1970)}]{Brinkman_1970}
\bibinfo{author}{\bibfnamefont{W.~F.} \bibnamefont{Brinkman}} \bibnamefont{and}
  \bibinfo{author}{\bibfnamefont{T.~M.} \bibnamefont{Rice}},
  \bibinfo{journal}{Phys. Rev. B} \textbf{\bibinfo{volume}{2}},
  \bibinfo{pages}{1324} (\bibinfo{year}{1970}),
  \urlprefix\url{https://link.aps.org/doi/10.1103/PhysRevB.2.1324}.

\bibitem[{\citenamefont{Trugman}(1988)}]{Trugman_1988}
\bibinfo{author}{\bibfnamefont{S.~A.} \bibnamefont{Trugman}},
  \bibinfo{journal}{Phys. Rev. B} \textbf{\bibinfo{volume}{37}},
  \bibinfo{pages}{1597} (\bibinfo{year}{1988}),
  \urlprefix\url{https://link.aps.org/doi/10.1103/PhysRevB.37.1597}.

\bibitem[{\citenamefont{Shraiman and Siggia}(1988)}]{Shraiman_1988}
\bibinfo{author}{\bibfnamefont{B.~I.} \bibnamefont{Shraiman}} \bibnamefont{and}
  \bibinfo{author}{\bibfnamefont{E.~D.} \bibnamefont{Siggia}},
  \bibinfo{journal}{Phys. Rev. Lett.} \textbf{\bibinfo{volume}{61}},
  \bibinfo{pages}{467} (\bibinfo{year}{1988}),
  \urlprefix\url{https://link.aps.org/doi/10.1103/PhysRevLett.61.467}.

\bibitem[{\citenamefont{Vidmar et~al.}(2015)\citenamefont{Vidmar, Ronzheimer,
  Schreiber, Braun, Hodgman, Langer, Heidrich-Meisner, Bloch, and
  Schneider}}]{Vidmar_2015}
\bibinfo{author}{\bibfnamefont{L.}~\bibnamefont{Vidmar}},
  \bibinfo{author}{\bibfnamefont{J.~P.} \bibnamefont{Ronzheimer}},
  \bibinfo{author}{\bibfnamefont{M.}~\bibnamefont{Schreiber}},
  \bibinfo{author}{\bibfnamefont{S.}~\bibnamefont{Braun}},
  \bibinfo{author}{\bibfnamefont{S.~S.} \bibnamefont{Hodgman}},
  \bibinfo{author}{\bibfnamefont{S.}~\bibnamefont{Langer}},
  \bibinfo{author}{\bibfnamefont{F.}~\bibnamefont{Heidrich-Meisner}},
  \bibinfo{author}{\bibfnamefont{I.}~\bibnamefont{Bloch}}, \bibnamefont{and}
  \bibinfo{author}{\bibfnamefont{U.}~\bibnamefont{Schneider}},
  \bibinfo{journal}{Phys. Rev. Lett.} \textbf{\bibinfo{volume}{115}},
  \bibinfo{pages}{175301} (\bibinfo{year}{2015}),
  \urlprefix\url{https://link.aps.org/doi/10.1103/PhysRevLett.115.175301}.

\bibitem[{\citenamefont{Ronzheimer et~al.}(2013)\citenamefont{Ronzheimer,
  Schreiber, Braun, Hodgman, Langer, McCulloch, Heidrich-Meisner, Bloch, and
  Schneider}}]{Bloch_2013}
\bibinfo{author}{\bibfnamefont{J.~P.} \bibnamefont{Ronzheimer}},
  \bibinfo{author}{\bibfnamefont{M.}~\bibnamefont{Schreiber}},
  \bibinfo{author}{\bibfnamefont{S.}~\bibnamefont{Braun}},
  \bibinfo{author}{\bibfnamefont{S.~S.} \bibnamefont{Hodgman}},
  \bibinfo{author}{\bibfnamefont{S.}~\bibnamefont{Langer}},
  \bibinfo{author}{\bibfnamefont{I.~P.} \bibnamefont{McCulloch}},
  \bibinfo{author}{\bibfnamefont{F.}~\bibnamefont{Heidrich-Meisner}},
  \bibinfo{author}{\bibfnamefont{I.}~\bibnamefont{Bloch}}, \bibnamefont{and}
  \bibinfo{author}{\bibfnamefont{U.}~\bibnamefont{Schneider}},
  \bibinfo{journal}{Phys. Rev. Lett.} \textbf{\bibinfo{volume}{110}},
  \bibinfo{pages}{205301} (\bibinfo{year}{2013}),
  \urlprefix\url{https://link.aps.org/doi/10.1103/PhysRevLett.110.205301}.

\bibitem[{\citenamefont{Lanczos}(1950)}]{lanczos}
\bibinfo{author}{\bibfnamefont{C.}~\bibnamefont{Lanczos}}, \bibinfo{journal}{J.
  Res. Nat. Bur. Stand.} \textbf{\bibinfo{volume}{45}}, \bibinfo{pages}{255}
  (\bibinfo{year}{1950}).

\bibitem[{\citenamefont{Jakli\v{c} and Prelov\v{s}ek}(2000)}]{jaklic2000}
\bibinfo{author}{\bibfnamefont{J.}~\bibnamefont{Jakli\v{c}}} \bibnamefont{and}
  \bibinfo{author}{\bibfnamefont{P.}~\bibnamefont{Prelov\v{s}ek}},
  \bibinfo{journal}{Advances in Physics} \textbf{\bibinfo{volume}{49}},
  \bibinfo{pages}{1} (\bibinfo{year}{2000}),
  \eprint{https://doi.org/10.1080/000187300243381},
  \urlprefix\url{https://doi.org/10.1080/000187300243381}.

\bibitem[{\citenamefont{Prelov{\v{s}}ek and
  Bon{\v{c}}a}(2013)}]{prelovsek_book}
\bibinfo{author}{\bibfnamefont{P.}~\bibnamefont{Prelov{\v{s}}ek}}
  \bibnamefont{and}
  \bibinfo{author}{\bibfnamefont{J.}~\bibnamefont{Bon{\v{c}}a}},
  \emph{\bibinfo{title}{Ground state and finite temperature Lanczos methods}}
  (\bibinfo{publisher}{Springer-Verlag Berlin Heidelberg},
  \bibinfo{year}{2013}), vol. \bibinfo{volume}{176} of
  \emph{\bibinfo{series}{Springer Series in Solid-State Sciences}},
  chap.~\bibinfo{chapter}{1}, pp. \bibinfo{pages}{1--30}.

\bibitem[{\citenamefont{Shastry and Sutherland}(1990)}]{shastry}
\bibinfo{author}{\bibfnamefont{B.~S.} \bibnamefont{Shastry}} \bibnamefont{and}
  \bibinfo{author}{\bibfnamefont{B.}~\bibnamefont{Sutherland}},
  \bibinfo{journal}{Phys. Rev. Lett.} \textbf{\bibinfo{volume}{65}},
  \bibinfo{pages}{243} (\bibinfo{year}{1990}).

\bibitem[{\citenamefont{Poilblanc}(1991)}]{Poilblanc_1991}
\bibinfo{author}{\bibfnamefont{D.}~\bibnamefont{Poilblanc}},
  \bibinfo{journal}{Phys. Rev. B} \textbf{\bibinfo{volume}{44}},
  \bibinfo{pages}{9562} (\bibinfo{year}{1991}),
  \urlprefix\url{https://link.aps.org/doi/10.1103/PhysRevB.44.9562}.

\bibitem[{\citenamefont{Bon\ifmmode~\check{c}\else \v{c}\fi{}a and
  Prelov\ifmmode~\check{s}\else \v{s}\fi{}ek}(2003)}]{Bonca_2003}
\bibinfo{author}{\bibfnamefont{J.}~\bibnamefont{Bon\ifmmode~\check{c}\else
  \v{c}\fi{}a}} \bibnamefont{and}
  \bibinfo{author}{\bibfnamefont{P.}~\bibnamefont{Prelov\ifmmode~\check{s}\else
  \v{s}\fi{}ek}}, \bibinfo{journal}{Phys. Rev. B}
  \textbf{\bibinfo{volume}{67}}, \bibinfo{pages}{085103}
  (\bibinfo{year}{2003}),
  \urlprefix\url{https://link.aps.org/doi/10.1103/PhysRevB.67.085103}.

\bibitem[{\citenamefont{Fehske et~al.}(1995)\citenamefont{Fehske, R\"oder,
  Wellein, and Mistriotis}}]{fehske1995}
\bibinfo{author}{\bibfnamefont{H.}~\bibnamefont{Fehske}},
  \bibinfo{author}{\bibfnamefont{H.}~\bibnamefont{R\"oder}},
  \bibinfo{author}{\bibfnamefont{G.}~\bibnamefont{Wellein}}, \bibnamefont{and}
  \bibinfo{author}{\bibfnamefont{A.}~\bibnamefont{Mistriotis}},
  \bibinfo{journal}{Phys. Rev. B} \textbf{\bibinfo{volume}{51}},
  \bibinfo{pages}{16582} (\bibinfo{year}{1995}),
  \urlprefix\url{https://link.aps.org/doi/10.1103/PhysRevB.51.16582}.

\bibitem[{\citenamefont{La~Magna and Pucci}(1997)}]{magna1997}
\bibinfo{author}{\bibfnamefont{A.}~\bibnamefont{La~Magna}} \bibnamefont{and}
  \bibinfo{author}{\bibfnamefont{R.}~\bibnamefont{Pucci}},
  \bibinfo{journal}{Phys. Rev. B} \textbf{\bibinfo{volume}{55}},
  \bibinfo{pages}{14886} (\bibinfo{year}{1997}),
  \urlprefix\url{https://link.aps.org/doi/10.1103/PhysRevB.55.14886}.

\bibitem[{\citenamefont{Bonca et~al.}(2000)\citenamefont{Bonca, Katrasnik, and
  Trugman}}]{bonca2000}
\bibinfo{author}{\bibfnamefont{J.}~\bibnamefont{Bonca}},
  \bibinfo{author}{\bibfnamefont{T.}~\bibnamefont{Katrasnik}},
  \bibnamefont{and} \bibinfo{author}{\bibfnamefont{S.~A.}
  \bibnamefont{Trugman}}, \bibinfo{journal}{Phys. Rev. Lett.}
  \textbf{\bibinfo{volume}{84}}, \bibinfo{pages}{3153} (\bibinfo{year}{2000}),
  \urlprefix\url{https://link.aps.org/doi/10.1103/PhysRevLett.84.3153}.

\bibitem[{\citenamefont{Alexandrov}(Springer, Dordrecht, 2007)}]{alexandrov07}
\bibinfo{author}{\bibfnamefont{A.~S.} \bibnamefont{Alexandrov}},
  \bibinfo{journal}{\textit{Polarons in Advanced Materials}, Springer Series in
  Material Sciences Vol.103}  (\bibinfo{year}{Springer, Dordrecht, 2007}).

\bibitem[{\citenamefont{Kornilovitch}(2002)}]{kornilovitch}
\bibinfo{author}{\bibfnamefont{P.~E.} \bibnamefont{Kornilovitch}},
  \bibinfo{journal}{EPL (Europhysics Letters)} \textbf{\bibinfo{volume}{59}},
  \bibinfo{pages}{735} (\bibinfo{year}{2002}),
  \urlprefix\url{http://stacks.iop.org/0295-5075/59/i=5/a=735}.

\bibitem[{\citenamefont{Salje et~al.}(1995)\citenamefont{Salje, Alexandrov, and
  Liang}}]{alex_book_1995}
\bibinfo{author}{\bibfnamefont{E.}~\bibnamefont{Salje}},
  \bibinfo{author}{\bibfnamefont{A.}~\bibnamefont{Alexandrov}},
  \bibnamefont{and} \bibinfo{author}{\bibfnamefont{W.}~\bibnamefont{Liang}},
  \emph{\bibinfo{title}{Polarons and Bipolarons in High-Tc Superconductors and
  Related Materials}} (\bibinfo{publisher}{Cambridge University Press},
  \bibinfo{year}{1995}).

\bibitem[{\citenamefont{Alexandrov and Mott}(1994)}]{alexandrov}
\bibinfo{author}{\bibfnamefont{A.}~\bibnamefont{Alexandrov}} \bibnamefont{and}
  \bibinfo{author}{\bibfnamefont{N.~F.} \bibnamefont{Mott}},
  \bibinfo{journal}{Rep. Prog. Phys.} \textbf{\bibinfo{volume}{57}},
  \bibinfo{pages}{1197} (\bibinfo{year}{1994}).

\bibitem[{\citenamefont{Alexandrov and Mott}(World Scientific, Singapore,
  1995)}]{alexandrov95}
\bibinfo{author}{\bibfnamefont{A.~S.} \bibnamefont{Alexandrov}}
  \bibnamefont{and} \bibinfo{author}{\bibfnamefont{N.~F.} \bibnamefont{Mott}},
  \bibinfo{journal}{Polarons and Bipolarons}  (\bibinfo{year}{World Scientific,
  Singapore, 1995}).

\end{thebibliography}

\end{document}